\documentclass[twocolumn]{aastex62}

\newcommand\harm{\textit{HARM${^2}$}}

\def\hii{H~\textsc{ii}}
\def\harm{\textit{HARM${^2}$}}
\def\orion{\textsc{Orion2}}

\def\msun{$\rm{M_{\rm \odot}}$}

\def\pap1{Paper~I}
\def\chandra{\textit{Chandra}}
\def\chianti{\textsc{Chianti}}

\def\ofb{\texttt{ROF$\mu_{\phi}$2}}
\def\ofsb{\texttt{ROF$\mu_{\phi}$20}}
\def\ofnob{\texttt{ROF$\mu_{\phi}\infty$}}
\def\wofb{\texttt{ROFW$\mu_{\phi}$2}}
\def\wofsb{\texttt{ROFW$\mu_{\phi}$20}}
\def\wofnob{\texttt{ROFW$\mu_{\phi}\infty$}}

\usepackage{amsmath}

\usepackage[normalem]{ulem}

\newcommand{\blue}[1]{{\textcolor{blue}{#1}}}

\usepackage{subfigure}

\usepackage{lineno}

\received{April 20, 2022}
\revised{October 30, 2022}
\accepted{October 31, 2022}
\submitjournal{ApJ}

\shorttitle{The Role of Stellar Feedback in Massive Star Formation}
\shortauthors{Rosen}

\begin{document}
\title{A Massive Star is Born: How Feedback from Stellar Winds, Radiation Pressure, and Collimated Outflows Limits Accretion onto Massive Stars}

\correspondingauthor{Anna Rosen}
\email{anna@ucsd.edu}

\author[0000-0003-4423-0660]{Anna L. Rosen}
\altaffiliation{ITC Fellow}
\altaffiliation{NSF Fellow}
\altaffiliation{UC Chancellor's Postdoctoral Fellow}
\affiliation{Center for Astrophysics $|$ Harvard \& Smithsonian, 60 Garden St, Cambridge, MA 02138, USA}
\affiliation{Center for Astronomy \& Space Sciences, University of California, San Diego, La Jolla, CA 92093, USA}


\begin{abstract}
Massive protostars attain high luminosities as they are actively accreting and the radiation pressure exerted on the gas in the star's atmosphere may launch isotropic high-velocity winds. These winds will collide with the surrounding gas producing shock-heated ($T\sim 10^7$ K) tenuous gas that adiabatically expands and pushes on the dense gas that may otherwise be accreted. We present a suite of 3D radiation-magnetohydrodynamic simulations of the collapse of massive prestellar cores and include radiative feedback from the stellar and dust-reprocessed radiation fields, collimated outflows, and, for the first time, isotropic stellar winds to model how these processes affect the formation of massive stars. We find that winds are initially launched when the massive protostar is still accreting and its wind properties evolve as the protostar contracts to the main-sequence. Wind feedback drives asymmetric adiabatic wind bubbles that have a bipolar morphology because the dense circumstellar material pinches the expansion of the hot shock-heated gas. We term this the ``wind tunnel effect." If the core is magnetized, wind feedback is less efficient at driving adiabatic wind bubbles initially because magnetic tension delays their growth. We find that wind feedback eventually quenches accretion onto $\sim$30~$\rm{M_{\rm \odot}}$ protostars that form from the collapse of the isolated cores simulated here. Hence, our results suggest that $\gtrsim$30~$\rm{M_{\rm \odot}}$ stars likely require larger-scale dynamical inflows from their host cloud to overcome wind feedback. 
Additionally, we discuss the implications of observing adiabatic wind bubbles with \chandra\ while the massive protostars are still highly embedded.

\end{abstract}

\keywords{methods: numerical --- stars: formation --- stars: massive --- turbulence}

\section{Introduction}
Massive stars ($\gtrsim 8$ \msun) directly influence their environment with their intense radiation fields, fast stellar winds, and supernova explosions at the end of their short lives. The energy and momentum imparted to the interstellar medium (ISM) by these processes, known as stellar feedback, leads to inefficient star formation in giant molecular clouds (GMCs) and their host galaxies \citep{Hopkins2014a, Gatto2017a, Krumholz2019a, Grudic2022a}. Stellar feedback has a direct consequence for the formation of massive stars because it can halt the accretion of material onto the star, thereby potentially affecting the maximum mass a star can achieve \citep{Olivier2021a}. 

Massive stars form from the gravitational collapse of dense ($\sim$$10^4-10^7 \; \rm cm^{-3}$), magnetized, turbulent, and cold ($\sim$$10$~K) molecular gas in GMCs \citep[see reviews by][]{Tan2014a, Rosen2020a}. Due to their short Kelvin-Helmholtz timescales, they achieve their main-sequence luminosities while they are actively accreting \citep{Palla1991a, Palla1992a,  Behrend2001a,  Hosokawa2009a}. Additionally, they produce powerful jets (hereafter collimated outflows) that are magnetically launched via the star-disk interaction \citep{Shu1988a, Pelletier1992a, Kolligan2018a}. Because of this, a common signature of massive star formation is the entrained molecular outflows that emanate from their birth sites when they are highly embedded and actively accreting \citep{Maud2015b, Rosen2020b, Avison2021a}. Significant theoretical attention has been given to the role that radiation pressure and collimated outflows play in massive star formation \citep[e.g.,][]{Krumholz2009a, Cunningham2011a, Kuiper2015a, Rosen2016a, Rosen2019a, Kee2019a, Rosen2020b}. However, no attention has been given to the role that stellar winds (i.e., the stellar surface material that is launched due to the radiation pressure exerted on the gas in the star's atmosphere) might play during their formation. Since massive protostars attain high luminosities while they are actively accreting and contracting to the main-sequence they should launch radiatively-driven stellar winds when they are relatively cool \citep[$T_{\rm eff} \gtrsim 15$~kK,][]{Vink2001a, Vink2018a}. Therefore, stellar wind feedback may be important in regulating accretion onto massive protostars earlier, as compared to other modes of feedback, as they contract to the main-sequence and are heavily embedded.


Stellar winds from main-sequence massive stars are roughly isotropic and leave the stellar surface at or above the escape speed of the star, which is $>$$10^3 \; \rm km\; s^{-1}$ for main-sequence massive stars \citep{Leitherer1992a, Vink2001a}. These fast winds collide with the surrounding ISM and thermalize producing hot, shocked stellar wind material \citep{Castor1975a, Weaver1977a, Koo1992a}. Assuming all of the wind kinetic energy injected (where $\dot{E} = 1/2 \dot{M}_{\rm w} v_{\rm w}^2$ is the rate of kinetic energy injected by winds, and $\dot{M}_{\rm w}$ and $v_{\rm w}$ are the wind mass-loss rate and wind velocity) is thermalized the resulting temperature of the hot shock-heated gas should attain temperatures of \citep{Rosen2021a}

\begin{equation}
T_{\rm X} \approx 10^7 \left( \frac{v_{\rm w}}{1000 \; \rm km \; s^{-1}}\right)^2 \;  \rm K.
\end{equation}
This hot gas will then cool via adiabatic expansion rather than significant radiative losses since cooling at these high temperatures is inefficient resulting in energy-driven (i.e., energy-conserving) stellar wind feedback \citep{Weaver1977a, Koo1992a, Rosen2014a, Rosen2021a}. The resulting expansion will sweep up a dense shell of entrained molecular material, producing energy-driven adiabatic wind bubbles. However, at the shell interface a significant fraction of the kinetic energy from stellar winds can be lost via radiative cooling because the hot shock-heated gas and cold and turbulent interstellar gas can mix to form $T\sim 10^4-10^5$~K gas that cools efficiently, thereby reducing the impact of wind feedback on larger scales \citep{Rosen2014a, Toala2018a, Lancaster2021a, Lancaster2021b}.

The hot gas produced by wind feedback emits thermal X-rays and therefore can be observed with X-ray telescopes like \chandra\ \citep{Lopez2011a, Rosen2014a}. \citet{Olivier2021a} performed the first multi-wavelength study of a large sample of galactic ultra-compact and compact \hii\ regions ($R_{\hii}\lesssim 0.5$~pc) that surround young massive stellar systems to determine the importance of different feedback mechanisms during their formation. They studied the hot $T\ge10^6-10^7$~K gas produced by stellar winds, the direct and dust-reprocessed radiation pressures, and the warm $T \approx 10^4$~K photoionized gas produced by photoionization. To study the importance of wind feedback they used \chandra\ archival data of 26 \hii\ regions, of which only 6 had reliable detections ($\ge 10$~photons). Given that these \hii\ regions were unresolved they were unable to separate the diffuse X-ray emission associated with the hot X-ray emitting gas produced by wind feedback and the stellar sources that typically have hard $\gtrsim 3$~keV emission. Therefore, they were only able to determine upper limits for the hot gas pressures, $P_{X}$, in their sample. Regardless, they found that the majority of \hii\ regions in their sample are dominated by the dust-reprocessed radiation pressure and they were unable to determine the importance of wind feedback. Therefore, it still remains uncertain how important wind feedback is during the early formation of massive stars. 

Numerical simulations have shed light on the importance of wind feedback from massive stars in the context of star cluster formation. \citet{Dale2014a} simulated the effect of photionization and momentum-conserving wind feedback (i.e., they neglected the thermalization of stellar winds) from massive stars with self-consistent star formation and found that the momentum injected by winds was dynamically unimportant. However, they were likely underestimating the effect of wind feedback because they did not include the kinetic energy injected by stellar winds and therefore neglected the hot, thermalized gas that should be produced by the shock-heating of stellar winds. 
\citet{Geen2021a} simulated the effect of photo-ionizing radiation and wind feedback on GMC scales for main-sequence massive stars, including both the momentum and energy injection by winds, and found that the adiabatic wind bubbles that form are initially confined and grow to have complex asymmetric morphologies. Likewise, \citet{Grudic2022a} performed the first star cluster formation simulation with self-consistent individual star formation and protostellar evolution with feedback from collimated outflows, radiation pressure, photoionization, and stellar winds (including both momentum and energy injection), with the new STARFORGE framework \citep{Grudic2021a}, to determine how these processes work in concert to quench star formation in GMCs. However, in their simulation stellar winds are only launched from massive stars once they reach the main-sequence. They found that feedback from radiation and winds are responsible for quenching star formation in GMCs. While these simulations demonstrated the importance of wind feedback from massive stars they did not study the direct impact wind feedback has on the formation of individual massive protostars and the resulting wind bubbles that may form while they are actively accreting and contracting to the main-sequence.

In this paper, we investigate these effects by performing 3D radiation-magnetohydrodynamics (RMHD) numerical simulations of the collapse of magnetized and unmagnetized turbulent massive prestellar cores into massive stellar systems, including both radiative, collimated outflow, and, for the first time, radiatively-driven isotropic wind feedback to explore how both the energy and momentum injected by stellar winds affects massive star formation. This paper is organized as follows: we describe our numerical methodology and simulation design in Section~\ref{sec:meth}. We present and discuss our results in Sections~\ref{sec:results} and \ref{sec:disc}, respectively. Finally, we conclude and summarize our results in Section~\ref{sec:conc}.

\section{Numerical Method}
\label{sec:meth}

In this paper, we simulate the formation of massive stars from the gravitational collapse of isolated magnetized and unmagnetized turbulent massive pre-stellar cores with the \orion\  adaptive mesh refinement (AMR) constrained-transport gravito-radiation-magnetohydrodynamics (RMHD) simulation code \citep{Orion2021}. \orion\ includes MHD \citep{Li2012a}, radiative transfer \citep{Krumholz2007a, Shestakov2008a, Rosen2017a}, self-gravity \citep{truelove1998a}, and Lagrangian accreting sink particles \citep{Krumholz2004a} that include a protostellar evolution model used to represent them as radiating (proto)stars \citep{Offner2009a}. The star particles are coupled to sub-grid prescriptions that models stellar feedback from both collimated protostellar outflows \citep{Cunningham2011a, Rosen2020a} and isotropic radiatively-driven stellar winds \citep{Offner2015a, Rosen2021a}. We describe the equations solved by \orion\ and the boundary conditions for the simulations in Section~\ref{sec:numerics}, the simulation initial conditions in Section~\ref{sec:ics}, our refinement and sink creation requirements in Section~\ref{sec:ref}, and the stellar radiation, outflow, and wind feedback prescriptions in Section~\ref{sec:feedback}.

\subsection{Evolution Equations and Boundary Conditions}
\label{sec:numerics}
The full gravito-RMHD equations solved by \orion\ that describe the dynamics of the fluid-sink (star) particle system for the simulations presented in this work are:

\begin{eqnarray}
\label{eqn:com}
\frac{\partial \rho}{\partial t} & = & -\mathbf{\nabla} \cdot \left( \rho \mathbf{v} \right) - \sum_{i}  \dot{M}_{a,i} W_{a}(\mathbf{x} - \mathbf{x}_i)  \nonumber\\
	                                   &&+ \sum_i \dot{M}_{o,i} W_{o,i}(\mathbf{x} - \mathbf{x}_i)  \nonumber\\
	                                   &&+ \sum_i \dot{M}_{w,i} W_{w,i}(\mathbf{x} - \mathbf{x}_i)
\\
\label{eqn:cop}
\frac{\partial \left(\rho \mathbf{v} \right)}{\partial t}  &=& -\mathbf{\nabla} \cdot \left( \rho \mathbf{v}  \bf{v} \right) - \mathbf{\nabla} \left( P + \frac{B^2}{8\pi}\right) + \frac{1}{4\pi} \mathbf{B} \cdot \mathbf{\nabla} \mathbf{B} \nonumber\\ 
	&& {} - \rho \mathbf{\nabla} \phi - \lambda \mathbf{\nabla} E_{\rm R}  - \sum_i \dot{\mathbf{p}}_{a,i}W_{a}(\mathbf{x} - \mathbf{x}_i) \nonumber\\
	&& + \sum_i \dot{\mathbf{p}}_{\rm rad, \it i} + \sum_i \dot{\mathbf{p}}_{o, \it i} W_{o,i}(\mathbf{x} - \mathbf{x}_i) \nonumber \\
	&& + \sum_i \dot{\mathbf{p}}_{w, \it i} W_{w,i}(\mathbf{x} - \mathbf{x}_i)
\\
\label{eqn:coe}
\frac{\partial \left( \rho e\right)}{\partial t} &=& - \mathbf{\nabla} \cdot \left[ (\rho e + P + \frac{B^2}{8 \pi})\textbf{v}   
	- \frac{1}{4\pi} \mathbf{B} (\bf{v} \cdot \mathbf{B}) \right]\nonumber \\
&& - \rho \mathbf{v} \cdot \mathbf{\nabla} \phi - \kappa_{\rm 0P} \rho(4\pi B_{P} - cE_{\rm R} ) \nonumber \\ 
	&&+ \lambda \left( 2 \frac{\kappa_{\rm 0P}}{\kappa_{\rm 0R}} - 1 \right) \mathbf{v} \cdot \mathbf{\nabla} E_{\rm R} \nonumber \\
	&& - \left(\frac{\rho}{m_{\rm p}} \right)^2 \Lambda(T_{\rm g}) -  \sum_i \dot{\varepsilon}_{a,i} W_{a,i}(\mathbf{x} - \mathbf{x}_i) \nonumber \\
	&&+ \sum_i \dot{\mathbf{\varepsilon}}_{\rm rad, \it i} + \sum_i \dot{\mathbf{\varepsilon}}_{o,i} W_{o,i}(\mathbf{x} - \mathbf{x}_i)  \nonumber \\
	&&+ \sum_i \dot{\mathbf{\varepsilon}}_{w,i} W_{w}(\mathbf{x} - \mathbf{x}_i)
\\
\label{eqn:coer}
\frac{\partial E_{\rm R}}{\partial t} &=& \mathbf{\nabla} \cdot \left( \frac{c \lambda}{\kappa_{\rm 0R} \rho} \mathbf{\nabla} E_{\rm R} \right) + \kappa_{\rm 0P} \rho \left(4\pi B_P - c E_{\rm R} \right) \nonumber \\
	&&- \lambda \left( 2 \frac{\kappa_{0 \rm P}}{\kappa_{0 R}} - 1\right) \mathbf{v} \cdot \nabla E_{\rm R} 
	- \nabla \cdot \left( \frac{3 - R_2}{2} \mathbf{v} E_{\rm R}\right) \nonumber \\ 
	& & +  \left(\frac{\rho}{m_{\rm p}} \right)^2 \Lambda(T_{\rm g}) 
\\
\label{eqn:induct}
\frac{\partial \bf{B}}{\partial t}&& = \mathbf{\nabla} \times (\bf{v} \times \bf{B})
\\
\label{eqn:mdot}
\frac{d M_{\rm i}}{dt} &&= \dot{M}_{a,i}  - \dot{M}_{o,i} - \dot{M}_{w,i}\\
\label{eqn:dvi}
\frac{d \bf{x}_{i}}{dt} &&=\frac{\bf{p}_{i}}{M_{\rm i}} \\
\label{eqn:dpi}
\frac{d\bf{p}_i}{dt} &&= -M_i \nabla \phi + \dot{\bf{p}}_{a,i} \\
\label{eqn:pois}
\nabla^2 \phi &&= 4 \pi G \left[ \rho + \sum_i M_i \delta(\bf{x} - \bf{x}_i)\right].
\end{eqnarray}

\noindent
In these equations, $\rho$ is the gas density, $\rho \mathbf{v}$  is the momentum density, $\rho e$ is the total internal plus kinetic gas energy density, $E_{\rm R}$ is the radiation energy density in the rest frame of the computational domain, $\mathbf{B}$ is the magnetic field, and $\phi$ is the gravitational potential. Equations \ref{eqn:com}-\ref{eqn:coer} describe conservation of gas mass, gas momentum, gas total energy, and radiation total energy, respectively. They include terms describing the exchange of these quantities with the star particles (which are denoted by the subscript $i$), and exchange of energy and momenta between radiation, magnetic fields, gas, and star particles including contributions from stellar radiation, collimated protostellar outflows, and isotropic stellar winds. Equation~\ref{eqn:induct} is the induction equation that describes the time evolution of the magnetic field in the ideal MHD limit, which assumes the magnetic field and fluid are well-coupled. \orion\ uses a constrained transport scheme that maintains $\mathbf{\nabla} \cdot \mathbf{B} = 0$ to machine accuracy \citep{Li2012a}.

The gas follows an ideal equation of state so that the gas pressure is defined as
\begin{equation}
\label{eqn:pres}
P=\frac{\rho k_{\rm B}T}{\mu m_{\rm H}} = \left( \gamma-1\right) \rho e_{\rm T},
\end{equation}
where $T$ is the gas temperature, $\mu$ is the mean molecular weight,  $\gamma$ is the ratio of specific heats, and $e_{\rm T}$ is the thermal energy of the gas per unit mass. We take $\mu=2.33$, which is appropriate for molecular gas of solar composition (i.e., the initial prestellar core material composition), and $\gamma=5/3$, which is appropriate for molecular gas at temperatures too low to excite the rotational levels of H$_2$ and the hot shock-heated gas produced by stellar wind feedback \citep{Weaver1977a, Rosen2016a, Rosen2021a}. Additionally, we assume the fluid is a mixture of dust and gas with a dust-to-gas mass ratio of 0.01 and assume the gas and dust temperatures are the same since, at the high densities modeled in this work, the dust will be thermally coupled to the gas \citep[][]{Hopkins2022a}.

\orion\ uses the (gray) flux limited diffusion (FLD) approximation to model the radiative emission and absorption by the gas and dust, which assumes that the radiative flux in the comoving frame is related to the gradient of the radiation energy density, to follow the evolution of the radiation field coupled to the fluid (see \citet{Krumholz2007a}  and \citet{Rosen2016a} for more detail). The radiation-specific quantities in Equations~\ref{eqn:coe}-\ref{eqn:coer} are the blackbody function $B_{\rm P} = c a_{\rm R} T^4/(4\pi)$, the density- and temperature-dependent Planck- and Rosseland-mean opacities $\kappa_{\rm 0P}$ and $\kappa_{\rm 0R}$ computed in the frame co-moving with the gas, the (dimensionless) flux limiter $\lambda$, and the Eddington factor $R_2$ \citep{Helling2000a, Semenov2003a, Krumholz2007a}. These last two quantities originate from the FLD approximation. Lastly, we include continuum and metal line cooling, which only becomes significant when $T \gtrsim 10^3$ K (i.e., when dust  begins to sublime), with the cooling function  $\Lambda(T)$and we assume $\mu=0.6$, which is appropriate for ionized gas of solar composition \citep{Cunningham2011a}. 

Equations~\ref{eqn:mdot}-\ref{eqn:dpi} describe the dynamical evolution of the (proto)star particles, as indexed by the subscript $i$, which accrete nearby gas and interact with the fluid via gravity, stellar radiation, collimated protostellar outflows, and isotropic stellar winds. We describe the modeling of their feedback (i.e., the momentum and energy injected into the fluid) associated with their radiation fields, outflows, and winds in Section~\ref{sec:feedback}, but note here that the radiation, outflows, and wind specific terms in Equations~\ref{eqn:com}-\ref{eqn:coe} affiliated with star particles are denoted with the $\mathrm{rad}$, $o$, and $w$ subscripts, respectively. The star particles are characterized by their mass $M_i$, position $\mathbf{x}_{i}$, momentum $\mathbf{p}_i$, angular momentum that describes the particle's spin axis $\mathbf{J}_i$, and luminosity ($\dot{\varepsilon}_{\rm{rad, \, i}}$), as determined by the protostellar evolution model described in \citet{Offner2009a}. They accrete mass, momentum, and energy from the computational grid via the weighting kernel $W_{a}(\mathbf{x}-\mathbf{x}_i)$, which is non-zero only within 4 radial cells of each particle following the sink particle accretion algorithm described in \citet{Krumholz2004a},  at rates $\dot{M}_{a,i}$, $\dot{\mathbf{p}}_{a,i}$, and $\dot{\varepsilon}_{a,i}$, respectively. The star particles' angular momentum and spin axis directions are updated via the subgrid model described in \citet{Fielding2015a}. Lastly, Equation~\ref{eqn:pois} describes how the gravitational potential of the gas is advanced and includes contributions from the fluid and sink (star) particles. 

The boundary conditions for the hydrodynamic, gravity, and radiation solvers are as follows. We impose outflow boundary conditions for the hydrodynamic update by setting the gradients of the hydrodynamic quantities $\left(\rho, \; \rho \bf{v}, \rho e \right)$ to be zero at the domain when advancing equations~\ref{eqn:com}-\ref{eqn:coe} \citep{Cunningham2011a, Myers2013a, Rosen2016a, Rosen2019a} and set the gravitational potential, $\phi$, to zero at all boundaries since the core boundaries are far removed from the domain boundaries when solving Equation~\ref{eqn:pois}. Finally, for each radiation update, we impose Marshak boundary conditions that bathe the simulation volume with a blackbody radiation field equal to $E_0 = 1.21 \times 10^{-9} \rm{\; erg \; cm^{-3}}$ corresponding to a 20 K blackbody but  allow for radiation generated within the simulation volume to escape freely \citep{Krumholz2009a, Cunningham2011a, Myers2013a, Rosen2016a, Rosen2019a}.

\begin{table*}
	\begin{center}
	\begin{tabular}{ l c  c  c  c  c  c  c c}
	\hline
	\textbf{Run} & & \ofnob & \wofnob\  & \ofb\ & \wofb\ & \ofsb\ & \wofsb\ \\
	\\
	\hline
	\textbf{Physical Parameter}\\
	\hline
	\hline
	Mass-to-flux ratio & $\mu_{\rm \phi}$ & $\infty$ & $\infty$ & 2 & 2 & 20 & 20\\
	Magnetic Field Strength [$\rm  mG $] & $B_z$  & 0 & 0 & 0.81 & 0.81 & 0.081 & 0.081\\
	Rad. Feedback? & & Yes & Yes  & Yes & Yes & Yes  & Yes\\
	Outflows? & & Yes & Yes & Yes & Yes & Yes & Yes  \\
	Winds? & & No & Yes & No & Yes & No & Yes  \\
	\hline
	\textbf{Simulation Outcome}\\
	\hline
	\hline
	Simulation end time [$t_{\rm ff}$]  & & 1.05  & 0.91 & 1.39 & 1.05 & 0.92 & 0.85 \\
	Massive star mass [\msun] & & 35.76 & 30.25 & 34.00 & 31.89 & 28.75 & 27.58 \\
	Number of sinks  & & 15 & 14 & 2 & 2 & 9 & 9 \\
	Star Formation Efficiency & $\frac{M_{\rm \star,  \;tot}}{M_{\rm c}}$ & 0.32 & 0.24 & 0.23 & 0.21 & 0.25 & 0.23 \\
	\hline 
	
	\\
	\end{tabular}
		\caption{
	\label{tab:sim}
Physical parameters and simulation outcomes for the simulations presented in this work. Each simulation begins with an isolated cold ($T_{\rm c}=20$~K) dusty molecular magnetized or unmagnetized pre-stellar core with mass $M_{\rm c} = 150~M_{\rm \odot}$ and radius $R_{\rm c}=0.1$~pc, corresponding to a surface density of $\Sigma=1$~$\rm g/cm^2$.  Each core is placed in a 0.4~pc box and is seeded with supersonic turbulence with an initial velocity dispersion of $\sigma_{\rm 1D}=1.2$~km/s. Each simulation has a base grid of $128^3$ cells and we allow for 4 levels of refinement corresponding to a maximum resolution of 40~au. 
}
	\end{center}
\end{table*}

\subsection{Initial Conditions}
\label{sec:ics}
In this work, we perform six simulations of the collapse of turbulent, massive prestellar cores with feedback from stellar radiation, collimated outflows, and isotropic stellar winds to determine how these feedback mechanisms affect the formation and mass growth of massive stars. The first three simulations do not include isotropic stellar wind feedback (runs \ofnob, \ofb, and \ofsb) and are only used to compare to identical simulations that do include wind feedback (runs \wofnob, \wofb, and \wofsb) to determine how wind feedback alters the accretion flow onto massive stars and affects the gas dynamics and gas structure near the star. In these simulations, the sub-grid model for stellar winds is turned on when the star reaches an effective temperature of 12.5~kK following the wind mass-loss rate formulae from \citet{Vink2001a} described in Section~\ref{sec:winds}. Runs \ofnob\ and \wofnob\ do not include magnetic fields whereas \ofb\ and \wofb\ follow the collapse of magnetized cores. Likewise, runs \ofsb\ and \wofsb\ follow the collapse of weakly magnetized cores and are used to compare with the results of the non-magnetic and magnetic core collapse simulations in Section~\ref{sec:bfields} since these simulations include a magnetic field strength much weaker than those observed in dense molecular gas \citep{Crutcher2012a, Hull2019a}. The initial conditions for the simulations described next are summarized in Table~\ref{tab:sim}.

For all simulations presented here, we begin with an isolated prestellar core of molecular gas and dust  (dust-to-gas ratio of 0.01) with mass $M_{\rm c}=150 \; M_{\rm \odot}$ and radius $R_{\rm c} = 0.1$ pc  corresponding to a surface density of $\Sigma = M_{\rm c}/\pi R^2_{\rm c} = 1 \; \rm{g \; cm^{-2}}$ and mean density $\bar{\rho} = 2.4 \times 10^{-18} \; \rm{g \; cm^{-3}}$ ($1.2 \times 10^6 \; \rm{H \; nuclei \; cm^{-3}}$) consistent with massive prestellar core densities and radii in extreme massive star forming environments \citep[e.g.,][]{Battersby2014a, Ginsburg2015a, Ginsburg2018a, Cao2019a, Li2020a}. The corresponding characteristic free-fall collapse time scale is $t_{\rm ff}\approx 42.8 \; \rm{kyr}$. The core has a $\rho(r) \propto r^{- 3/2}$  density profile in agreement with observations of massive cores at the $\sim$0.1 pc scale and clumps at the $\sim$1 pc scale that find values of $\kappa_{\rm \rho} = 1.5-2$ \citep[e.g.,][]{Caselli1995a, Beuther2002b, Mueller2002a, Beuther2007a, Zhang2009a, Longmore2011a, Butler2012a, Battersby2014a, Stutz2016a, Beuther2021a}. The core's initial gas temperature is set to 20 K. Each core is placed in the center of the domain and the rest of the computational domain is filled with hot, diffuse gas with density $\rho_{\rm amb} = 0.01 \rho_{\rm edge}$ where $\rho_{\rm edge}$ is the density at the edge of the core and temperature $T_{\rm amb} = 2000$ K so that the core is in thermal pressure balance with the ambient medium. The opacity of the ambient medium is set to zero. 

Runs \ofb, \wofb,  \ofsb, and \wofsb\ include magnetic fields to determine how magnetic pressure and magnetic tension affect the accretion of material onto massive (proto)stars and the development of wind-driven bubbles produced by their wind feedback (when winds are included). For runs \ofb\ and \wofb, the initial magnetic field is initially uniform in the z direction with $\mathbf{B} = B_0 \hat{z}$ where $B_0= 0.81$ mG corresponding to a mass-to-flux ratio $\mu_\Phi = M_{\rm c}/M_{\rm \phi} \simeq 2 \pi G^{1/2} M_{\rm c}/ \Phi = 2$, where $\Phi = \pi R^2_c B_0$ is the magnetic flux through the core, consistent with observed values of $\Phi \simeq$ 2-3 \citep{Crutcher2012a}. For runs \ofsb\ and \wofsb\ we set $\mu_\Phi =20$, yielding an initial weak magnetic field strength of $B_0= 0.081$ mG. These weak-field magnetic runs are used for comparing the magnetic and non-magnetic simulations.

Observed massive prestellar cores and clumps contain supersonic turbulence \citep[e.g., see reviews by][]{Hull2019a, Rosen2020a}. Following this, turbulent motions for the cores modeled here are included by seeding the initial gas velocities ($v_x$, $v_y$, and $v_z$) with a velocity power spectrum $P(k) \propto k^{-2}$, with modes between $k_{\rm min} = 1$ to $k_{\rm max} = 256$, as expected for supersonic turbulence \citep{Padoan1999a, Boldyrev2002a, Cho2003a, Kowal2007a}. The turbulence mixture of gas is 1/3 compressive and 2/3 solenoidal, consistent with the natural mixture of a 3D fluid \citep{Kowal2007a, Kowal2010a}. The onset of turbulence modifies the density and magnetic field distribution. All simulations are initialized with the same velocity perturbation power spectrum and a velocity dispersion of  $\sigma_{\rm 1D}= 1.2$ km/s corresponding to $\alpha_{\rm vir} = 5 \sigma_{\rm 1D}^2 R_{\rm c}/G M_{\rm c}=1.1$ so that the core is roughly virialized \citep{Bertoldi1992a}.The turbulence will decay freely, however this simplification should have little effect on the results since the decay timescale, $\sim 2 R_{\rm c}/\sigma_{\rm 1D} \sim$~0.16 Myr \citep{Goldreich1995a}, is much longer than the runtime for the simulations presented in this work. 

We note that runs \ofb, \wofb,  \ofsb, and \wofsb\ also include magnetic pressure ($P_{\rm B} = B^2/8\pi$), yielding $\alpha_{\rm vir} = 5R_{\rm c}/G M_{\rm c} \left(\frac{1}{6} v_{\rm A}^2 + \sigma_{1D}^2 \right)$ where $v_{\rm A} = B/\sqrt{4\pi \rho}$ is the Alfven velocity. Therefore, the cores that include magnetic fields have slightly higher $\alpha_{\rm vir}$ with values of 1.4 and 1.12 for the initial field strengths of $B_{\rm z} = 0.81$~mG and $B_{\rm z} = 0.081$~mG, respectively. As shown in \citet{Rosen2020b}, this additional pressure term slows down the gravitational collapse of the pre-stellar core resulting in lower accretion rates onto the massive star and broader entrained molecular outflows that are eventually ejected from the core because the core material is less bound.

\subsection{Refinement and Sink Particle Creation Criteria}
\label{sec:ref}
Each simulation has a base grid with volume (0.4~pc)$^3$ discretized by $128^3$ cells and allows for four levels of refinement, resulting in a maximum resolution of 40~au. We note that the simulations presented in this work do not have the same maximum resolution of those presented in \citet{Rosen2020b}, which had a maximum level of resolution of 20~au, corresponding to 5 levels of refinement. Refinement up to 4 AMR levels was chosen for the simulations presented in this work because the shock-heated gas ($T \sim 10^7$~K) produced by wind feedback (i.e., the thermalization of the wind kinetic energy) causes the time step to drop significantly since the Courant condition used to calculate the time step, which is a requirement for numerical stability, depends on the gas sound speed and velocities \citep{Courant1967a}.

As the simulation evolves, the AMR algorithm automatically adds and removes finer grids. Cells are refined if they meet at least one of the following criteria: (1) any cell on level 0 (the base level) that has $\rho \ge \rho_{\rm edge}$, so that the entire core is refined to level 1; (2) any cell where the density in the cell exceeds the Jeans density given by
\begin{equation}
\label{eqn:rhoj}
\rho_{\rm max,J} = \frac{\pi J^2_{\rm max} c_{\rm s}^2}{G \Delta x^2_l} \left( 1 + \frac{0.74}{\beta^2}\right),
\end{equation}
\noindent
where $c_{s}=\sqrt{kT/\mu m_{\rm{p}}}$ is the isothermal sound speed, $\Delta x_l$ is the cell size on level $l$, $\beta=8\pi \rho c_s^2/B^2$ is the plasma parameter (i.e., the ratio of the thermal gas pressure to the magnetic pressure)\footnote{In the limit that as $B \rightarrow 0$ we have that $\beta \rightarrow \infty$ and Equation~\ref{eqn:rhoj} results to the classical Jeans limit \citep{Truelove1997a}.}, and $J_{\rm max}$ is the maximum allowed number of Jeans lengths per cell, which is set to 1/8 following the MHD Truelove Criterion \citep{Myers2013a};  (3) any cell that is located within at least sixteen cells of a sink particle; and (4) any cell within which the radiation energy density gradient exceeds $\nabla E_{\rm R} > 0.15 E_{\rm R}/\Delta x_{l}$. 

Star particles form on the maximum AMR level when the Jeans condition for a Jeans number of $N_J$ = 0.25 is exceeded following the resolution tests of \citet{Truelove1997a}. Star particles merge when they  pass within one accretion radius of each other if the smaller particle has a mass less than $0.04 \; M_{\rm \odot}$, corresponding to the threshold for the largest plausible mass at which second collapse occurs for the protostar.  Below this mass limit, a protostar represents a hydrostatic core that is several au in size and will likely be accreted by the more massive star \citep{Masunaga1998a, Masunaga2000a}. However, for masses above this value the protostar will have collapsed down to sizes of roughly several $R_{\rm \odot}$ and therefore it is not appropriate to assume if the protostar will accrete onto the nearby (proto)star since the accretion radius for a sink particle is $4 \times \Delta x_{4} = 160$~au.

\subsection{Stellar Feedback Modeling}
\label{sec:feedback}
The simulations presented in this work follow the same stellar radiation and collimated outflow feedback modeling described in \citet{Rosen2016a} and \citet{Rosen2020b}, with the addition of radiatively driven isotropic wind feedback described in \citet{Rosen2021a}. Each star particle has a (direct) isotropic stellar radiation field and collimated outflows, which are launched at the poles along the star's angular momentum axis, that inject energy ($\varepsilon$) and momentum ($\bf{p}$) into the surrounding fluid. Additionally, once the stellar effective temperature reaches $T_{\rm eff}=12.5$kK the star is hot and luminous enough to produce an isotropic radiatively driven wind \citep{Vink2001a}. We note that the outflows and winds are injected into the computational domain after the star accretes material and the resulting mass-loss is subtracted from the stellar mass before the stellar radius and luminosity are updated by the protostellar evolution model. To trace the outflow and wind material we add two passively advected scalars to represent the outflow and wind gas that is injected, respectively. The modeling of these feedback processes are summarized next. 

\subsubsection{Stellar Radiation}
\label{sec:rad}
The radiation pressure and radiative heating are modeled with the multi-frequency Hybrid Adaptive Ray-moment Method (HARM$^2$) presented in \citet{Rosen2016a} and \citet{Rosen2017a}, which treats both the direct (stellar) and indirect (dust-reprocessed) radiation fields. This method includes the direct solution of the frequency-dependent radiative transfer equation of the stellar radiation field along long characteristics (i.e., adaptive ray tracing) that are launched from the star isotropically and includes contributions from the stellar luminosity ($L_{\rm \star}$) and accretion luminosity given by
\begin{equation}
\label{eqn:Lacc}
L_{\rm acc} = f_{\rm rad} \frac{G M_{\rm \star} \dot{M_{\rm \star}}}{R_{\rm \star}},
\end{equation}  
\noindent 
where  $f_{\rm rad}= 3/4$ is the fraction of the gravitational potential energy of the accretion flow that is converted to radiation following \citet{Offner2009a}, $M_{\rm \star}$ is the stellar mass, and $R_{\rm \star}$ is the stellar radius determined by the sub-grid protostellar evolution model. The accretion luminosity is modeled as a blackbody spectrum with temperature $T_{\rm acc}=(L_{\rm acc}/(4\pi R_{\rm \star}^2 \sigma) )^{1/4}$ such that $L_{\rm acc} = \int^\infty_0 L_{\rm{acc,\nu}} d\nu$.  The rate of momentum and energy absorbed by the dusty fluid from the stellar radiation fields are given by $\dot{\bf{p}}_{\rm{rad}, i}$ and $\dot{\varepsilon}_{\rm{rad}, i}$ in Equations~\ref{eqn:cop} and \ref{eqn:coe}, respectively. The frequency-dependent stellar spectra and dust opacities are taken from \citet{Lejeune1997a} and \citet{Weingartner2001a} and are divided into ten frequency bins \citep[see Figure~1 of ][]{Rosen2016a}. The frequency range used for the stellar spectra, accretion spectra, and dust opacities is $1.87 \times 10^{12} - 3.29 \times 10^{16}$~$s^{-1}$, which covers the far-IR to EUV spectral range. Since dust is the primary absorber of the stellar radiation, dust absorption doesn't occur when the fluid temperature is $\ge1500$~K corresponding to the dust sublimation temperature. Instead the gas opacity is set to 0.01 $\rm cm^{2} \; g$, causing the warm gas carved out by feedback near the star to be effectively transparent to the stellar radiation field. This method is coupled to the gray FLD method described above to treat the (indirect) radiation field produced by thermal emission from dust \citep{Krumholz2007a, Rosen2016a, Rosen2017a}. 

We note that our method is similar to the radiative transfer hybrid method developed by \citet{Mignon-Risse2020a}, which uses the gray M1 closure relation for stellar irradiation and gray FLD for the dust and gas. However, the \harm\ multi-frequency approach for stellar irradiation used in this work more accurately captures the momentum and energy injection from stellar irradiation because the dust opacity increases by several orders of magnitude across the far-IR to EUV spectral range \citep[e.g., see Figure~1 of ][]{Rosen2016a}.

\subsubsection{Collimated Outflows}
\label{sec:outflows}
The magnetically launched collimated outflows are included as a sub-grid model first introduced by \citet{Cunningham2011a} and updated by \citet{Rosen2020b} since proper modeling of the launching of outflows requires sufficiently high resolution and non-ideal MHD processes \citep[e.g., sub-au scales, see][]{Kolligan2018a}, which is prohibitively expensive for the simulations presented in this work. In this sub-grid prescription, outflows are launched along the star's spin axis at the stellar poles, following the protostellar outflow model of \citet{Matzner2000a} that includes a collimation angle, $\theta_c$, and launching fraction, $f_{\rm w}$, that is related to the accretion rate. For the simulations presented here, we take $\theta_c=0.01$ and $f_{o}=0.21$, which assumes that 21\% of the accreted material is lost to outflows (i.e., $\dot{M}_{o,i} = f_{o} \dot{M}_{a,i}$). The outflows are injected in the eight nearest cells to the star (in radius) with the weighting kernel $W_{o,i}(\mathbf{x} - \mathbf{x}_i)$ described in \citet{Cunningham2011a}. 

The outflows are launched at a fraction $f_{k}=0.3$ of the Keplerian velocity, such that the outflow velocity is $v_0 = f_{k} \sqrt{GM_{\rm \star}/{R_{\rm \star}}}$. The outflows inject mass $\dot{M}_o$, momentum $\dot{p_o} =\dot{M}_0 v_o$, kinetic energy $\dot{E}_{k,o}=\frac{1}{2} \dot{M}_o v_o^2$, and thermal energy $\dot{E}_{T,o} = \frac{\dot{M}_o k T_o}{\mu_o m_{H} (\gamma-1)}$ where $T_o$ is the outflow gas temperature taken to be the star's effective temperature when $T_{\rm eff} < 10^4$~K and  $T_o = 10^4$~K otherwise and $\mu_o = 1.27$. The values used in these simulations for $f_o$  and $v_o$ are chosen to match the observed outflow momentum, $0.01\lesssim f_{o}v_{o} \lesssim 0.15$, in low- and high-mass star forming regions \citep{Cunningham2011a}. In agreement, \citet{Rosen2020b} found that these outflow properties agree well with the energetics from entrained molecular outflows observed in high-mass star forming regions \citep{Maud2015b}.

\subsubsection{Isotropic Stellar Winds}
\label{sec:winds}
In runs \wofnob, \wofb, and \wofsb, the star produces an isotropic stellar wind that is launched by the star's radiation pressure (i.e., a radiatively driven wind) when the star's effective temperature is $T_{\rm eff} \ge 12,500$~K following the mass-loss rate analytic formulae developed by \citet{Vink2001a}. As shown in Figure~\ref{fig:fb_props_mstar} and described in Section~\ref{sec:starprops}, wind launching typically starts when the contracting protostar in the simulations presented here reaches $\sim 20$~\msun. The wind mass-loss rate formulae used in these simulations, which depend on the stellar properties, are adapted from Monte Carlo simulations that follow the fate of a large number of photons from below the stellar photosphere that calculates the radiative acceleration (i.e., launching) of wind material from the stellar surface.  \citet{Vink2001a} show that the wind mass-loss rates experience a jump around $T_{\rm eff} \approx$ 25,000~K, known as the bi-stability jump, due to a change in the ionization state in the lower stellar photosphere (i.e., Fe iv recombines to form Fe iii) leading to Fe ions that are more efficient line drivers on the hot-side of the bi-stability jump. On the hot side of the bi-stability jump the mass-loss rate decreases by a factor of $\sim5$ but the wind velocity, $v_{\infty}$, increases by a factor of 2. 

Assuming solar metallicity, the wind mass-loss rate for the cool side of the bi-stability jump is
\begin{equation}
\label{eqn:mdot1}
\begin{split}
\log_{10}(\dot{M}_{\rm \star}) = &-6.688 + 2.210 \log_{10} \left(\frac{L_{\rm \star}}{10^5 \, L_{\rm \odot}} \right)\\
					       & - 1.339 \log_{10}\left( \frac{M_{\rm \star}}{30 \, M_{\rm \odot}}\right) \\
					      &  - 1.601 \log_{10} \left( \frac{v_{\rm \infty}/v_{\rm esc}}{2.0} \right) \\
					     &+ 1.07 \log_{10} \left( \frac{T_{\rm eff}}{20 \, \rm kK}\right)
\end{split}
\end{equation}
\noindent
where $v_{\rm esc} = \sqrt{2G M_{\rm \star}/R_{\rm \star}}$ is the escape speed at the star's surface.
On the hot-side of the bi-stability jump the wind mass-loss rate becomes: 
\begin{equation}
\label{eqn:mdot2}
\begin{split}
\log_{10}(\dot{M}_{\rm \star}) = &-6.697 + 2.194 \log_{10} \left(\frac{L_{\rm \star}}{10^5 \, L_{\rm \odot}} \right)\\
					       & - 1.313 \log_{10}\left( \frac{M_{\rm \star}}{30 \, M_{\rm \odot}}\right) \\
					        &- 1.226 \log_{10} \left( \frac{v_{\rm \infty}/v_{\rm esc}}{2.0} \right) \\
				&+ 0.933 \log_{10} \left( \frac{T_{\rm eff}}{40 \, \rm kK}\right) \\
				& - 10.92 \left( \log_{10} \left( \frac{T_{\rm eff}}{40 \, \rm kK}\right)  \right)^2.
\end{split}
\end{equation}
\noindent
To determine, which mass-loss recipe to use we first compute the bi-stability jump temperature following Equation~15 from \citet{Vink2001a}. If the (proto)star is on the cool (hot) side of the bi-stability the wind velocity ($v_{\rm \infty}$) is taken to be $1.3 \, v_{\rm esc}$ ($2.6 \, v_{\rm esc}$) following values of $v_{\infty}/v_{\rm esc}$ determined by both theory and observations of winds from B and O stars \citep[][and references therein]{Vink2001a}.

The winds are injected in a sphere encompassing the eight nearest cells to the star in radius ($R_{\rm w, \; inj} = 320$~au) with the weighting kernel $W_{w,i}(\mathbf{x} - \mathbf{x}_i)$ where each cell has an equal weight so that the total mass, momentum, and kinetic and thermal energies injected within this region are $\dot{M}_{w}$, $\dot{p}_w = \dot{M}_{w} v_{\rm \infty}$, $\dot{E}_{k,w}=\frac{1}{2} \dot{M}_w v_{\infty}^2$, and $\dot{E}_{T,w} = \frac{\dot{M}_w k T_w}{\mu_o m_{H} (\gamma-1)}$, respectively. We take $T_w = 10^4$~K and $\mu_o=1.27$.
 
\citet{Pittard2021a} found that in order to resolve the wind bubble dynamics accurately the wind injection radius, $R_{\rm{w, \, inj}}$, must be smaller than a characteristic injection radius given by
\begin{equation}
R_{\rm max, \, inj} = \left(\frac{\dot{M}_{\rm w} v_{\rm w}}{4\pi P_{\rm amb}}\right)^{1/2}
\end{equation}
\noindent
where $P_{\rm amb }$ is the total ambient pressure including the thermal, turbulent, and magnetic pressures. They find that the wind bubble momentum is within 25\% of the true value if  $\chi_{\rm w} \equiv R_{\rm{w, \, inj}}/R_{\rm{max, \, inj}} \le 0.1$. As will be shown in Figure~\ref{fig:fb_props_mstar}, the initial wind properties are $\dot{M}_{\rm w} \sim 10^{-7}$~\msun\ $\rm yr^{-1}$ and $v_{\rm w} \sim 500$~$\rm{km/s}$. For the core properties simulated in this work we find that our wind injection region is properly resolved with $\chi_{\rm w}$ values of 0.062, 0.076, and 0.063 for runs \wofnob, \wofb, and \wofsb, respectively.

\section{Results}
\label{sec:results}

\begin{figure*}
\centerline{\includegraphics[trim=0.8cm 14.5cm 0.2cm 0.2cm,clip,width=1\textwidth]{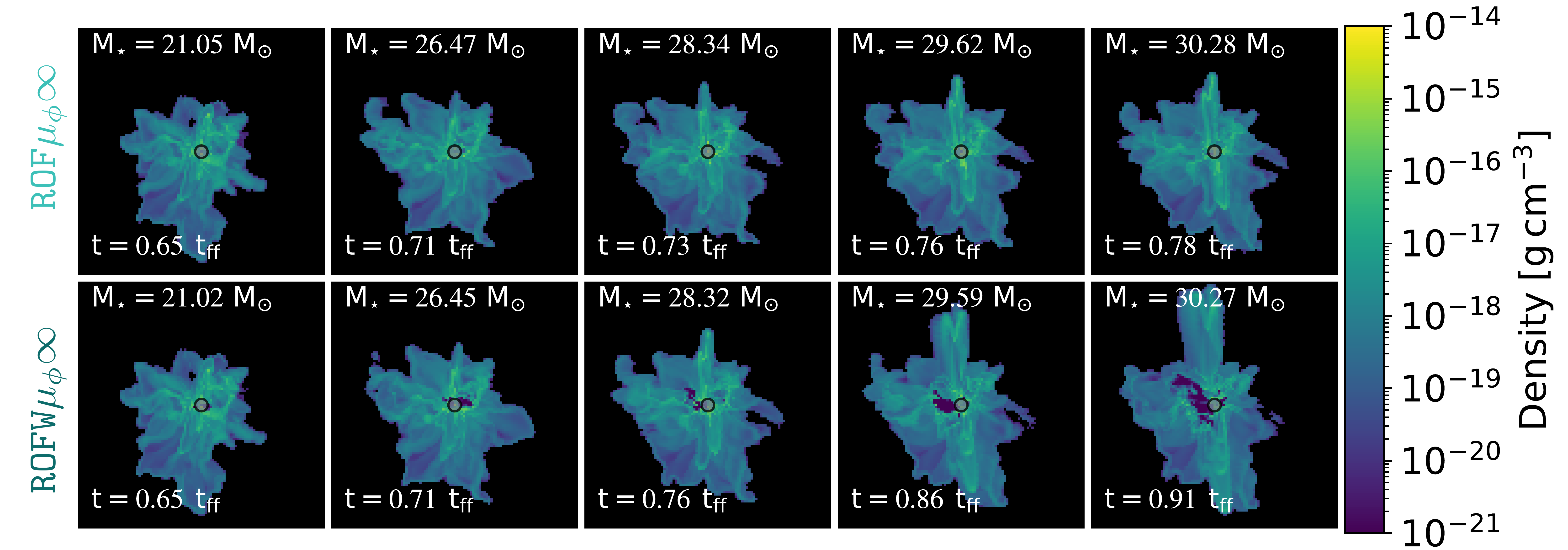}}
\caption{
\label{fig:densitySlices_NoB}
Density slices for runs \ofnob\ (top row) and \wofnob\ (bottom row) as a function of primary stellar mass. The most massive star (denoted by the gray circle) is located at the center of each panel and the slice is oriented such that its angular momentum axis points up, in order to highlight the density structure of the outflows. The primary stellar mass and the simulation time, in units of $t_{\rm ff}$, are shown in the bottom and top left corners of each panel, respectively. Each panel is (0.4 pc)$^2$.
}
\end{figure*}

\begin{figure*}
\centerline{\includegraphics[trim=0.8cm 14.5cm 0.2cm 0.2cm,clip,width=1\textwidth]{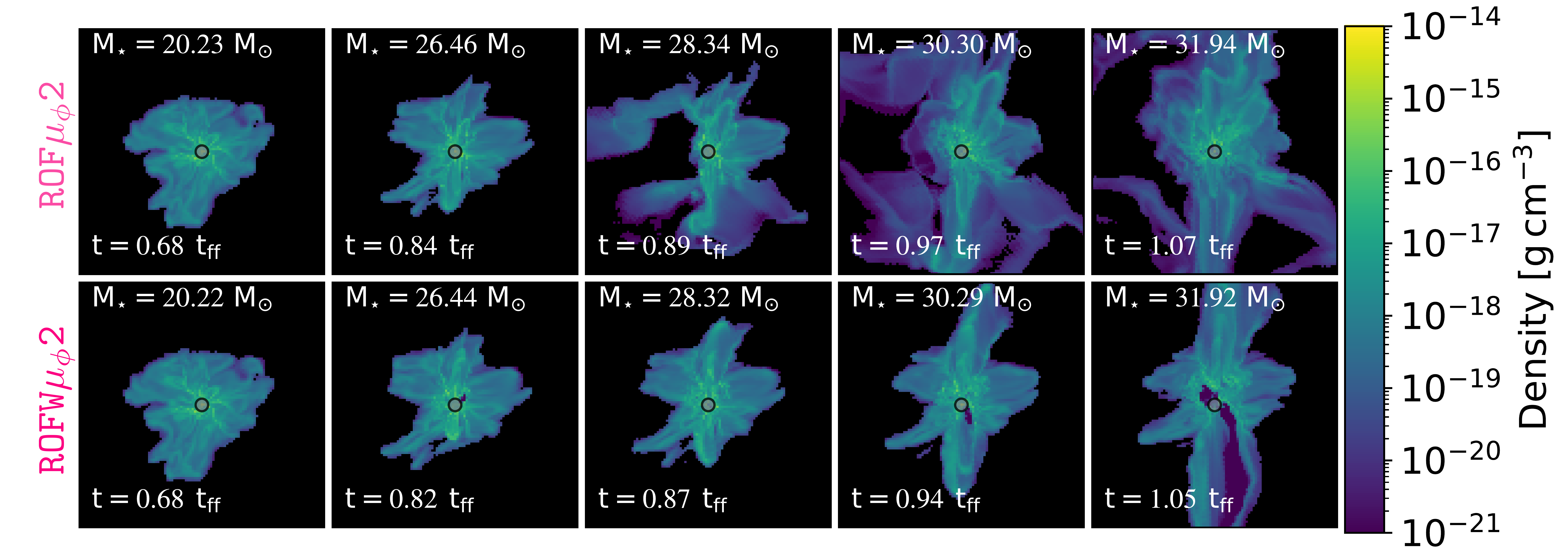}}
\caption{
\label{fig:densitySlices_B}
Same as Figure~\ref{fig:densitySlices_NoB} except now for runs \ofb\ (top row) and \wofb\ (bottom row).
}
\end{figure*}

\subsection{Density Structure}
\label{sec:density}
Figure~\ref{fig:densitySlices_NoB} shows density slice snapshots for runs \ofnob\ (top row) and \wofnob\ (bottom row) at similar primary (most massive) stellar mass. Each panel covers the full domain to show the density structure of the entire core and the entrained molecular outflows that are powered by the protostellar collimated outflows, which are eventually ejected from the core \citep[e.g., see][]{Rosen2020b}. We only include snapshots for each run when the primary star is hot enough to launch winds (i.e., when $T_{\rm eff} \gtrsim 12.5$~kK corresponding to when $M_{\rm \star} \gtrsim 21 \, M_{\rm \odot}$) to follow how wind feedback alters the density structure of gas near the primary star. Comparison of these snapshots for runs \ofnob\ (top row) and \wofnob\ (bottom row) show that inclusion of wind feedback leads to low density regions near the star that are not spherical even though the winds are launched isotropically. 

As we show in Section~\ref{sec:bubble}, this low-density gas is produced by the shock heating of the fast flowing stellar wind material \citep[e.g., see Section~2 of][]{Rosen2021a} that then undergoes adiabatic expansion (i.e., $P \, dV$ work) and carves out regions of low-density hot tenuous gas near the star. Since the density distribution near the primary star is turbulent with varying density, the asymmetry of the wind-driven ``bubble" demonstrates that the low-density gas produced by stellar wind feedback follows the path of least resistance (i.e, undergoes greater expansion in regions of lower density), thereby carving out a non-spherical structure near the star that grows in time as the winds are continuously injected. 

\begin{figure*}
\centerline{\includegraphics[trim=0.8cm 14.5cm 0.2cm 0.2cm,clip,width=1\textwidth]{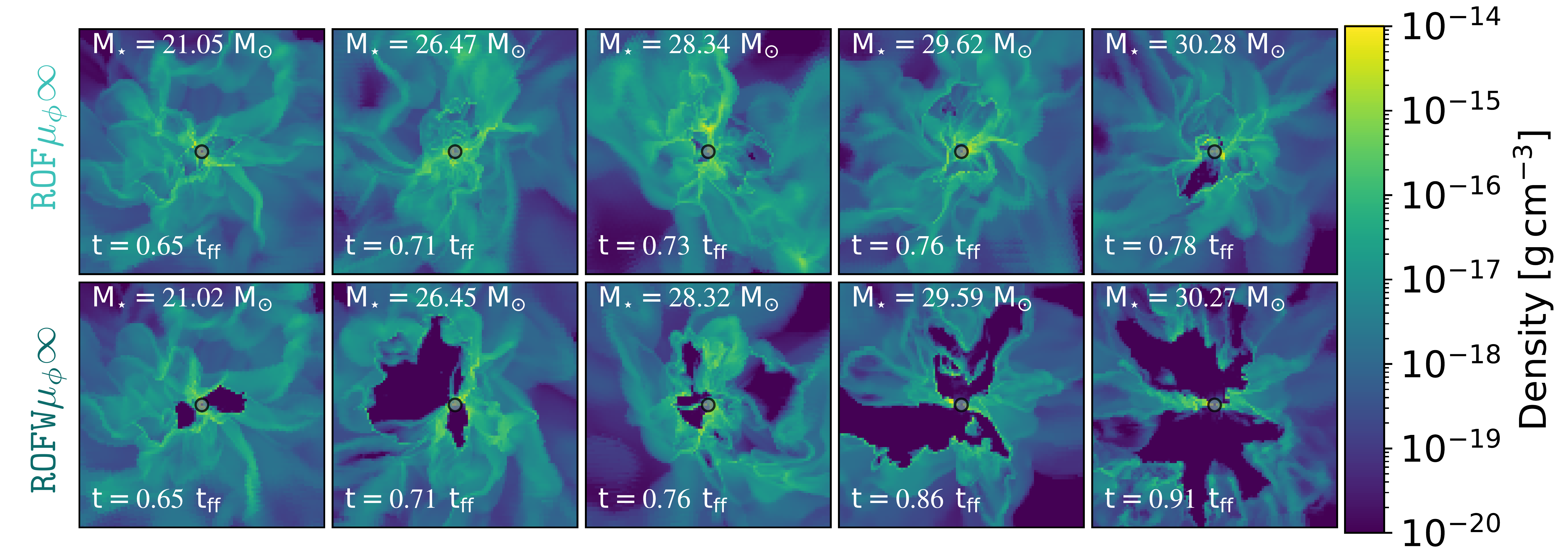}}
\caption{
\label{fig:densitySlices_NoBzoom}
Zoom-in density slices for runs \ofnob\ (top row) and \wofnob\ (bottom row) as a function of primary stellar mass. The most massive star is located at the center of each panel as marked by the gray circle. Each slice is oriented such that the mass-weighted angular momentum axis of the gas within a radius of 500~au from the primary star points up in order to highlight the wind-driven bubbles that are perpendicular to the dense circumstellar gas. The primary stellar mass and the simulations time are shown in the bottom and top left corners of each panel, respectively. Each panel is (0.1 pc)$^2$.
}
\end{figure*}

\begin{figure*}
\centerline{\includegraphics[trim=0.8cm 14.5cm 0.2cm 0.2cm,clip,width=1\textwidth]{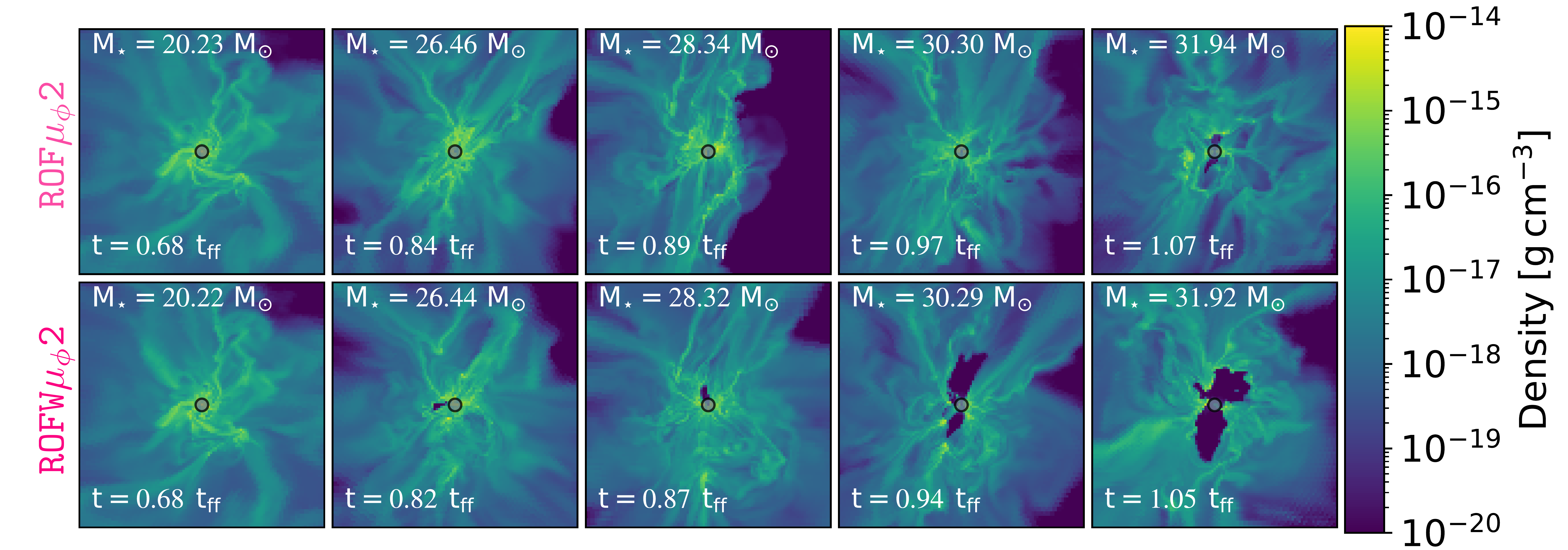}}
\caption{
\label{fig:densitySlices_Bzoom}
Same as Figure~\ref{fig:densitySlices_NoBzoom} except now for runs \ofb\ (top row) and \wofb\ (bottom row).
}
\end{figure*}

Comparison of runs \ofnob\ and \wofnob\ shows that inclusion of wind feedback slows down the mass growth of the primary star. Hence, the entrained outflows that are ejected from the core are larger and more pronounced by the end of run \wofnob\ as compared to run \ofnob\ at the same primary stellar mass because the outflows are injected for a longer period of time. Therefore, these results suggest that, in the absence of magnetic fields, wind feedback eventually reduces the accretion rate onto massive stars once the star has strong winds because the kinetic energy injected by stellar winds is thermalized near the star and generates hot, low-density gas that adiabatically expands pushing high density material away that may otherwise be accreted onto the star. We describe the evolving wind properties in detail in Section~\ref{sec:windprops}.

This picture changes when magnetic fields are included. Figure~\ref{fig:densitySlices_B} shows snapshots of the density slices for runs \ofb\ (top row) and \wofb\ (bottom row) at the same primary stellar mass. In run \wofb\ winds are launched when the star reaches $\sim 20 \, M_{\rm \odot}$, which is slightly less than the primary stellar mass in run \wofnob. This difference is due to the different accretion history of the primary star, which affects the radial evolution of the protostar and is described in more detail in Section~\ref{sec:starprops}. Comparison of runs \wofb\ and \wofnob\ in the bottom rows of Figures~\ref{fig:densitySlices_NoB} and \ref{fig:densitySlices_B} show that wind feedback is less effective at producing expanding adiabatic wind bubbles when the surrounding material is magnetized. For example, in run \wofnob\ wind feedback produces a small wind-driven bubble almost immediately once winds are launched but when the core is magnetized wind feedback doesn't produce the hot shock-heated, low density gas until the star reaches $\sim30 \; M_{\rm \odot}$. Furthermore, the last panel of the bottom row in Figure~\ref{fig:densitySlices_B} shows that the hot gas vents through the low density gas carved out by protostellar outflows in the bottom outflow lobe, thereby suggesting that the hot, low density gas produced by wind feedback more readily expands in the low density regions carved out by protostellar outflows. This is explored in more detail in Section~\ref{sec:entrained}.

Comparison of runs \ofb\ and \wofb\ suggest that the growth rate of the massive star does not change significantly when feedback from stellar winds are included if the core is magnetized and is described in more detail in Section~\ref{sec:bfields}. Regardless, these simulations demonstrate that feedback from stellar winds eventually produces localized low-density non-spherical cavities near the star that expand whereas protostellar outflows leads to larger scale collimated entrained molecular outflows that are eventually ejected from the core.

\subsection{Wind-Driven Bubble Morphology}
\label{sec:bubble}

\begin{figure*}
\centerline{\includegraphics[trim=0.2cm 0.2cm 0.2cm 0.2cm,clip,width=1\textwidth]{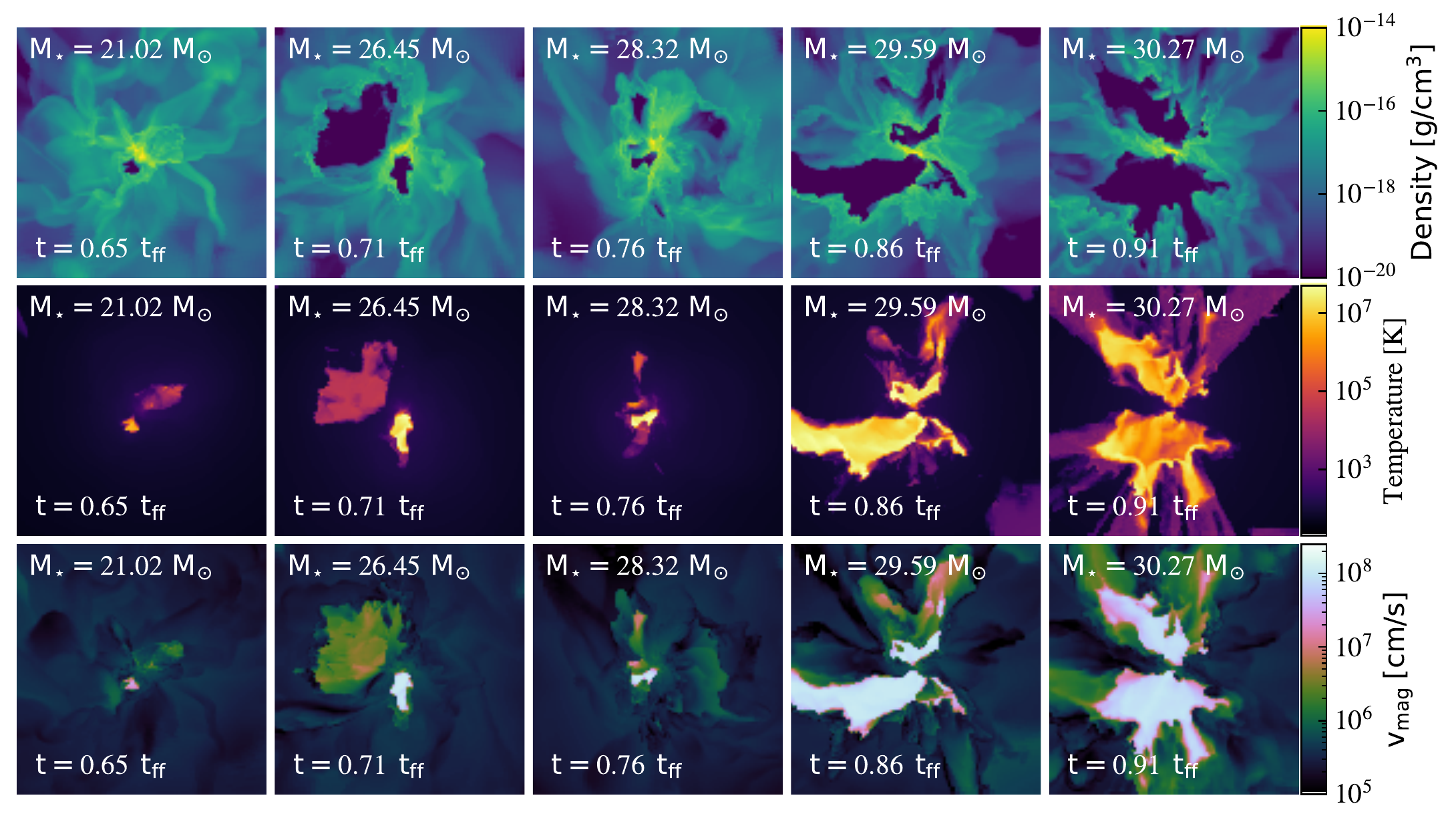}}
\caption{
\label{fig:bubble_noB}
Zoom-in thin density (top row), temperature (middle row), and velocity (bottom row) projections for run \wofnob\ as a function of primary stellar mass. The most massive star is located at the center of each panel and is marked by the gray circle. Each projection is oriented such that the mass-weighted angular momentum axis of the gas within a radius of 500~au from the primary star points up in order to highlight the wind-driven bubbles. 
Each panel is (0.1 pc)$^2$.
}
\end{figure*} 

\begin{figure*}
\centerline{\includegraphics[trim=0.2cm 0.2cm 0.2cm 0.2cm,clip,width=1\textwidth]{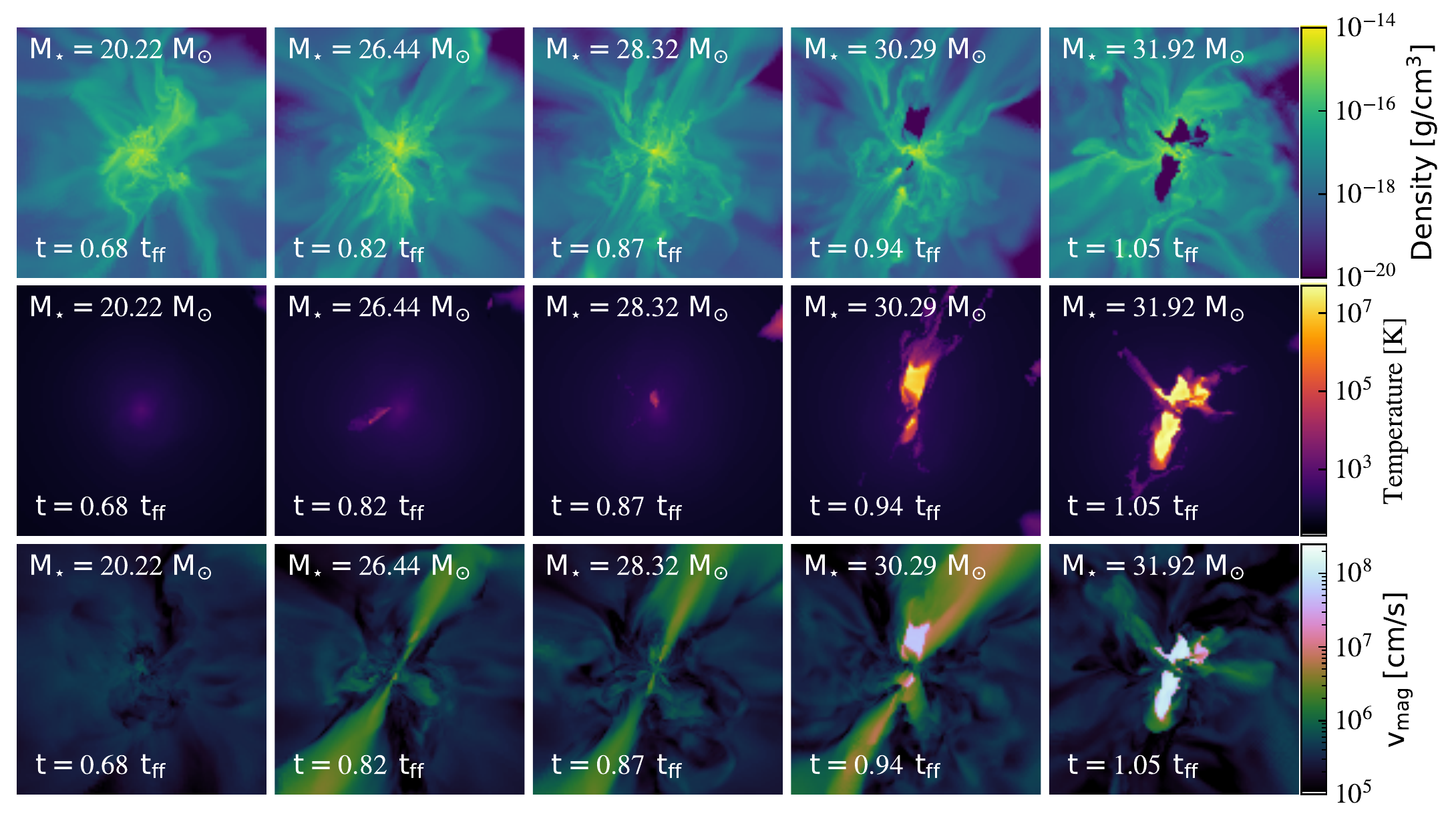}}
\caption{
\label{fig:bubble_B}
Same as Figure~\ref{fig:bubble_noB} except now for run \wofb.
}
\end{figure*}

Figures~\ref{fig:densitySlices_NoBzoom} and \ref{fig:densitySlices_Bzoom} show zoom-ins of the density slices presented in Figures~\ref{fig:densitySlices_NoB} and \ref{fig:densitySlices_B}, respectively; except now each slice is oriented such that the mass-weighted angular momentum axis of the gas near the massive star points up. This orientation is chosen to highlight the morphology of the low-density wind-driven bubbles or lobes that are launched perpendicular to the star's equatorial plane because along this direction the circumstellar gas is densest due to the conservation of angular momentum from the infalling core material, which is described in more detail in Section~\ref{sec:disk}. Comparison of runs \wofnob\ and \wofb\ (bottom rows) with the simulations that do not include wind feedback (runs \ofnob\ and \ofb; top rows) in Figures~\ref{fig:densitySlices_NoBzoom} and \ref{fig:densitySlices_Bzoom} demonstrate that the low-density bubbles produced near the primary star are due to wind feedback and not due to radiation pressure. 
The snapshots for the simulations that do not include wind feedback show that radiation pressure only begins to drive low-density radiation-pressure-dominated bubbles near the star when it reaches $\sim$$30 \; M_{\rm \odot}$ for run \ofnob\ and $\sim$$32 \; M_{\rm \odot}$ for run \ofb. Hence, our results suggest that wind feedback drives low density cavities before radiation pressure becomes strong enough to produce radiation-pressure-dominated bubbles.

These pinched low-density wind bubbles are produced by the shock-heating (i.e., thermalization) of the wind material as shown in Figures~\ref{fig:bubble_noB} and \ref{fig:bubble_B}  for runs \wofnob\ and  \wofb, respectively. These figures show zoom-ins of the thin mass-weighted projections of the gas density (top panels), gas temperature (middle panels), and gas velocity magnitude (bottom panels). We find that the shock-heated gas reaches temperatures of $T \sim$few$\times10^6-10^7$~K, corresponding to $c_{\rm s}\sim100-400 \; \rm km \, s^{-1}$, that adiabatically expands. However, the majority of the wind bubble volume contains free-flowing wind material with velocities  $\sim 10^3 \; \rm km \, s^{-1}$ that has yet to be thermalized but the velocities reach the expected shock-heated velocities near the dense shell interfaces. For run \wofnob\ we find that the wind-driven lobes can be crushed by the surrounding dense and infalling material. Eventually, as seen in the last panels of Figure~\ref{fig:bubble_noB} the winds become powerful enough to produce sustained wind-driven lobes. 

Comparison of Figures~\ref{fig:bubble_noB} and \ref{fig:bubble_B} show that the wind-driven bubbles are larger for the unmagnetized core and by the end of both runs the asymmetric lobes extend above and below the massive star. The wind-driven lobes by the end of run \wofb\ are smaller in size because they are confined by the magnetic tension in the surrounding dense gas, which is explored in more detail in Section~\ref{sec:bfields}. This confining effect was also found for the radiation-pressure-dominated bubbles presented in \citet{Rosen2020b}. 

\subsection{Accretion Disk Formation and Evolution}
\label{sec:disk}
\begin{figure*}
\centerline{\includegraphics[trim=1.45cm 16.5cm 0.2cm 0.2cm,clip,width=1\textwidth]{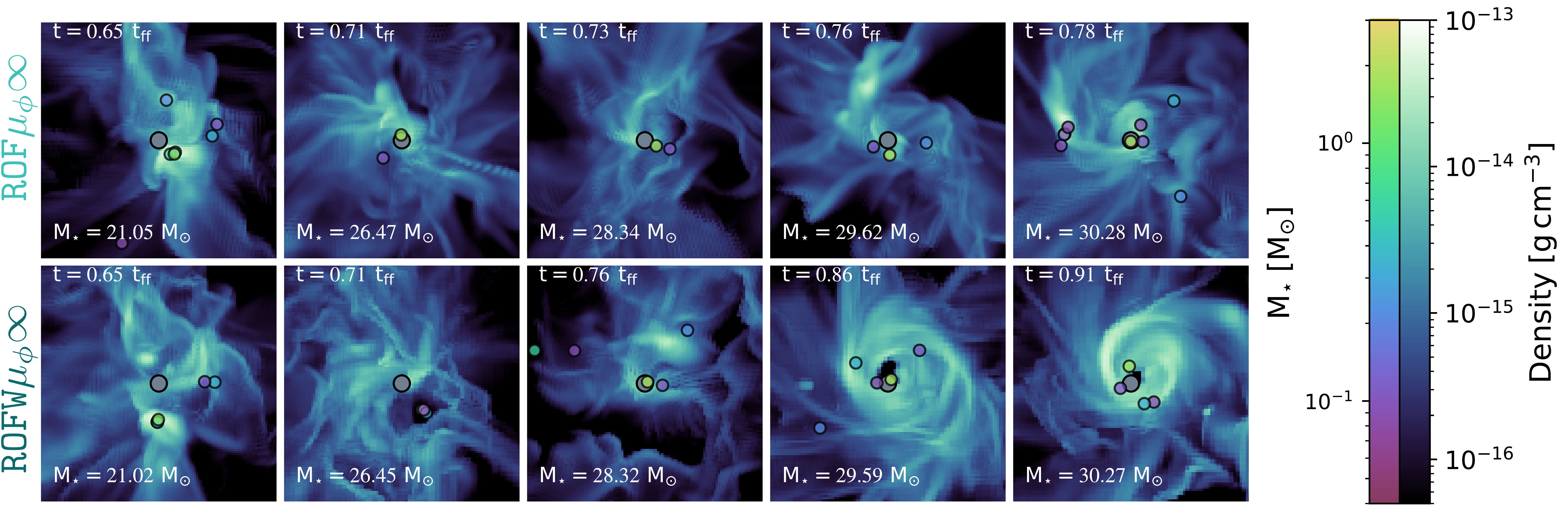}}
\caption{
\label{fig:disks_NoB}
Density-weighted density thin projections for runs \ofnob\ (top row) and \wofnob\ (bottom row) shown at similar primary stellar masses. The projections have been oriented so that the angular momentum axis of the material within 500 au of the primary star points out of the page, in order to highlight the accretion disk. In each panel, the most massive star is at the center (gray circle) and companion stars with masses greater than 0.04 $M_{\rm \odot}$ are over-plotted on all panels. The color of the star indicates its mass, as shown in the colorbar. Each panel is (3000 au)$^2$ in area, and the projection is taken over a depth of 500~au in front of and behind the massive star. 
}
\end{figure*}

Figure~\ref{fig:disks_NoB} shows thin density projections of the dense circumstellar material that surrounds the primary star along its equatorial plane for runs \ofnob\ (top row) and \wofnob\ (bottom row) at similar primary stellar mass. Eventually, a high-density circumstellar accretion disk (i.e., a resolved accretion disk with a radius larger than the 160 au accretion zone radius of the sink particle) forms at late times for both runs because the accretion disk structure depends on the angular momentum content of the collapsing core, which is larger for material that is farther out. Given that run \wofnob\  ran for a longer time the accretion disk is larger and more pronounced when the star reaches $\sim29$~$M_{\rm \odot}$ (i.e., the last two panels in the bottom row) as compared to run \ofnob\ at the same primary stellar mass. 

Comparison with the bottom row of Figure~\ref{fig:densitySlices_NoBzoom} shows that the presence of this high-density accretion disk causes the hot shock-heated gas produced by wind feedback to expand more freely along the polar directions of the primary star where the gas density is lower. This effect is the most pronounced for the last two snapshots in Figure~\ref{fig:densitySlices_NoBzoom} confirming that the dense accretion disk quenches the destructive effect of wind feedback near the star along the plane of the accretion disk. 

\begin{figure*}
\centerline{\includegraphics[trim=1.45cm 16.5cm 0.2cm 0.2cm,clip,width=1\textwidth]{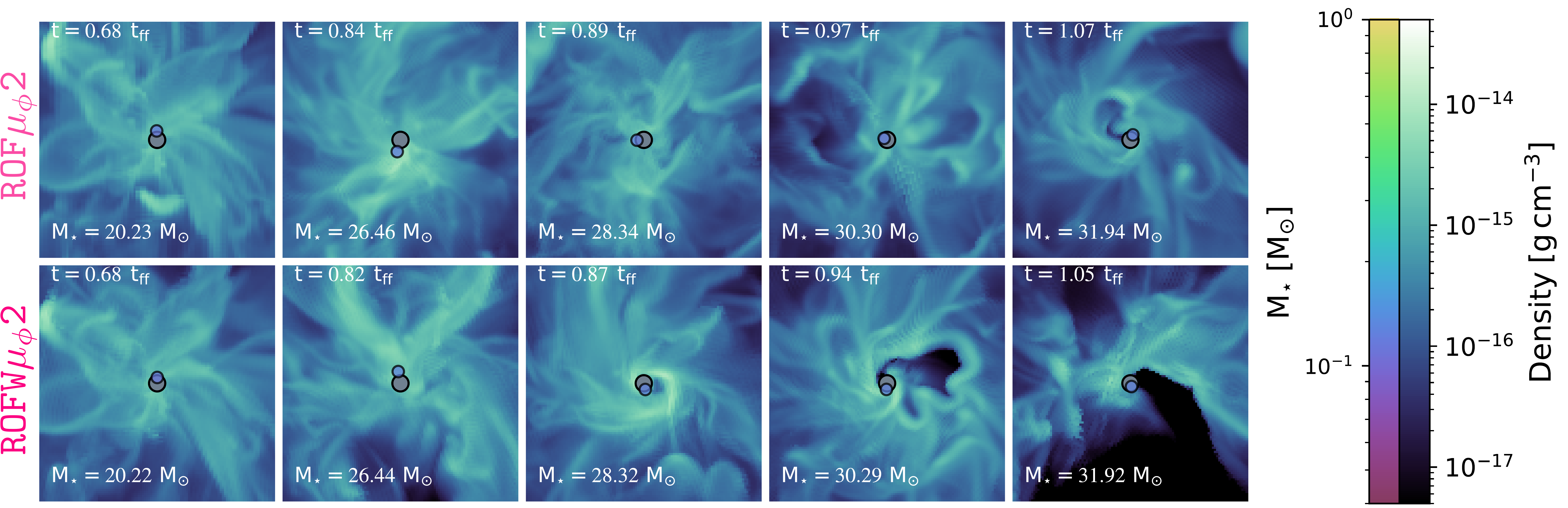}}
\caption{
\label{fig:disks_B}
Same as Figure~\ref{fig:disks_NoB} except now for runs \ofb\ (top row) and \wofb\ (bottom row).
}
\end{figure*}

In contrast, the influence of magnetic fields reduces disk formation due to magnetic braking as shown in Figure~\ref{fig:disks_B}, which shows thin density projections of the dense circumstellar material, as a function of primary stellar mass, that surround the primary star along its equatorial plane for runs \ofb\ (top row) and \wofb\ (bottom row). \citet{Rosen2020b} showed that magnetic braking, which removes angular momentum from the infalling material  as the core collapses inhibits the formation of a discernible accretion disk around the massive star, however higher resolution and/or non-ideal MHD effects such as ambipolar diffusion and Ohmic resistivity may reduce how much angular momentum is removed leading to smaller accretion disks than those produced when magnetic fields are not included \citep[e.g., ][]{Seifried2012b, Myers2013a, Zhao2020a, Mignon-Risse2021a, Commercon2021a}. Such effects are not explored in this work. Regardless, the bottom row of Figure~\ref{fig:disks_B} demonstrates that a noticeable accretion disk does not form around the primary star but the material near the primary star achieves high densities ($\rho \sim 10^{-15}$~$\rm g/cm^{3}$), which inhibits wind-bubble expansion leading to the bipolar morphology of the wind bubble. 

Eventually, as seen in the last two panels in the bottom row of Figure~\ref{fig:disks_B}, feedback from winds does blow away material near the star along the equatorial plane. However, when comparing to the last two panels of the bottom row of Figure~\ref{fig:densitySlices_Bzoom}, the hot gas is beamed along directions perpendicular to the dense circumstellar gas producing  wind-driven lobes above and below the star. These lobes are less pronounced than those produced in run \wofnob. This effect is likely due to accumulation of dense material along the equatorial plane near the star and the magnetic tension along the wind-driven lobe edges, which we explore in Section~\ref{sec:bfields}. 

\begin{figure*}
\centerline{\includegraphics[trim=0.3cm 0.2cm 0.2cm 0.2cm,clip,width=1\textwidth]{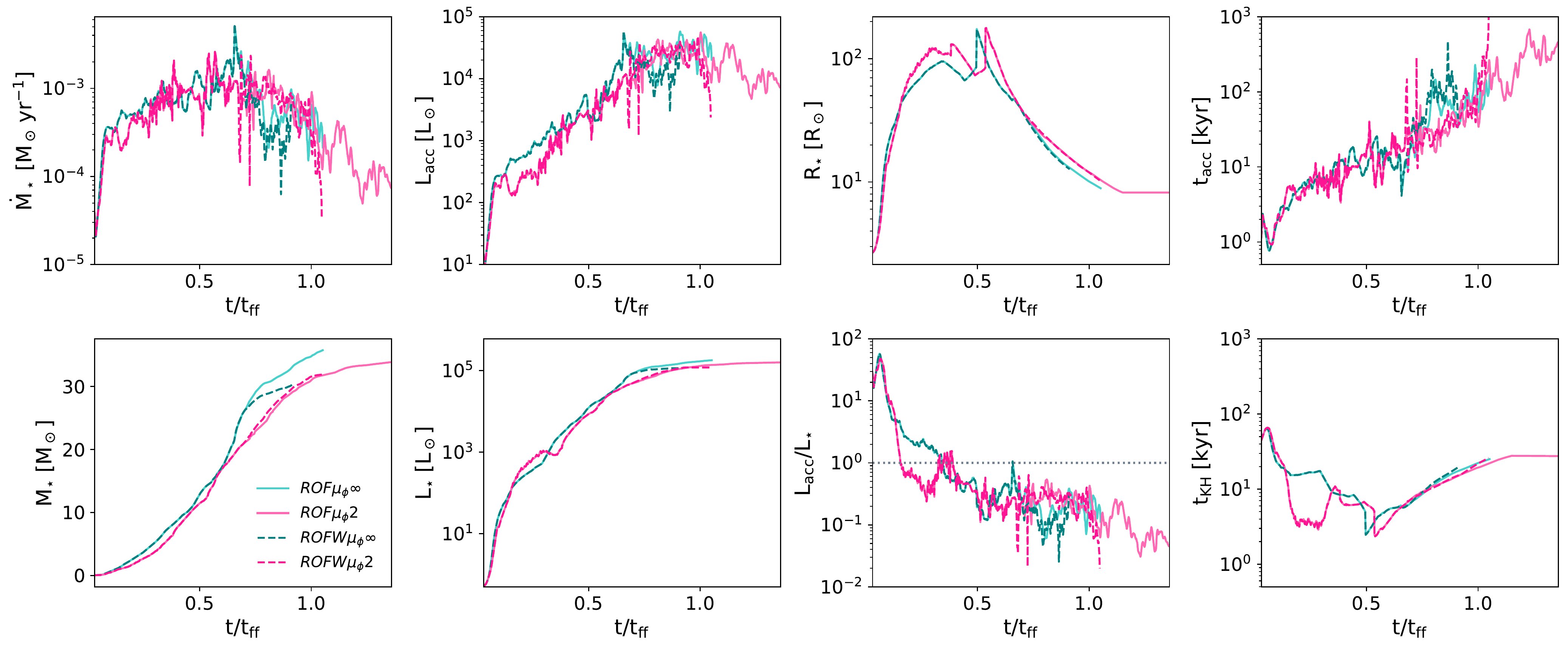}}
\caption{
\label{fig:star_props}
Primary protostar properties as a function of simulation time for runs  \ofnob\ (light teal solid lines), \wofnob\ (teal dashed lines), \ofb\ (light pink solid lines), and \wofb\ (dark pink dashed lines). The top row shows the primary star's accretion rate (left panel), accretion luminosity (middle left panel), radius (middle right panel), and accretion timescale $t_{\rm acc} = M_{\rm \star}/\dot{M}_{\rm \star}$ (right panel). The bottom row shows the primary star's mass (left panel), internal (stellar) luminosity (middle left panel), the ratio of the accretion to internal luminosity (middle right panel), and Kelvin Helmholtz timescale $t_{\rm KH} = G M_{\rm \star}^2/R_{\rm \star} L_{\rm \star}$ (right panel). The gray dashed line in the bottom middle right panel denotes where this $L_{\rm acc}/L_{\rm \star}$ is equal to 1.
}
\end{figure*}

\subsection{Massive Protostar Properties}
\label{sec:starprops}
The physical properties of the massive (primary) protostar as a function of simulation time for runs \ofnob, \ofb, \wofnob, and \wofb\ are shown in Figure~\ref{fig:star_props}. The left-hand column shows the accretion rate (top panel) and stellar mass (bottom panel). When magnetic fields are not included winds reduce the mass growth of the primary star: the accretion rate eventually drops when wind feedback becomes important and the mass growth of the massive star decreases appreciably when the star reaches $M_{\rm \star} \sim 27 \; M_{\rm \odot}$. Comparing these panels with the gas properties near the star in Figure~\ref{fig:bubble_noB} shows that this drop in the accretion rate is due to the development and subsequent growth of the wind-driven bubbles, thereby eventually quenching the accretion flow. The mass growth eventually plateaus suggesting that stellar winds may be responsible for halting accretion onto massive stars. 

Runs \ofb\ and \wofb\ show a different scenario. If the protostellar core is magnetized, the accretion rate actually increases for run \wofb\ once winds are launched by the star causing the mass growth of the massive star to increase slightly as compared to run \ofb. However, this increase is temporary because once the star reaches $\sim30 \; M_{\rm \odot}$ wind feedback begins to produce hot shock-heated gas near the star that gradually expands as shown in the last two columns in Figure~\ref{fig:bubble_B}. As will be shown in the next subsection, this behavior occurs when the  protostar crosses the bi-stability jump, described in Section~\ref{sec:winds}, where its wind velocity increases by a factor of $\sim2$. This increase in velocity will reduce radiative losses for the shock-heated gas during the free-expansion stage of the wind bubble (i.e., the ``fast winds" scenario by \citet{Koo1992a}) thereby making wind feedback more effective. Hence, when winds are initially launched the magnetized material near the primary star delays the effect of wind feedback in reducing the accretion flow but this is short lived and eventually, at a greater stellar mass than seen in run \wofnob\ ($\sim31$~\msun\ versus $\sim27$~\msun), wind feedback reduces accretion onto the massive star. This is discussed in more detail in Section~\ref{sec:bfields}.

\begin{figure}
\centerline{\includegraphics[trim=0.2cm 0.2cm 0.2cm 0.2cm,clip,width=1.0\columnwidth]{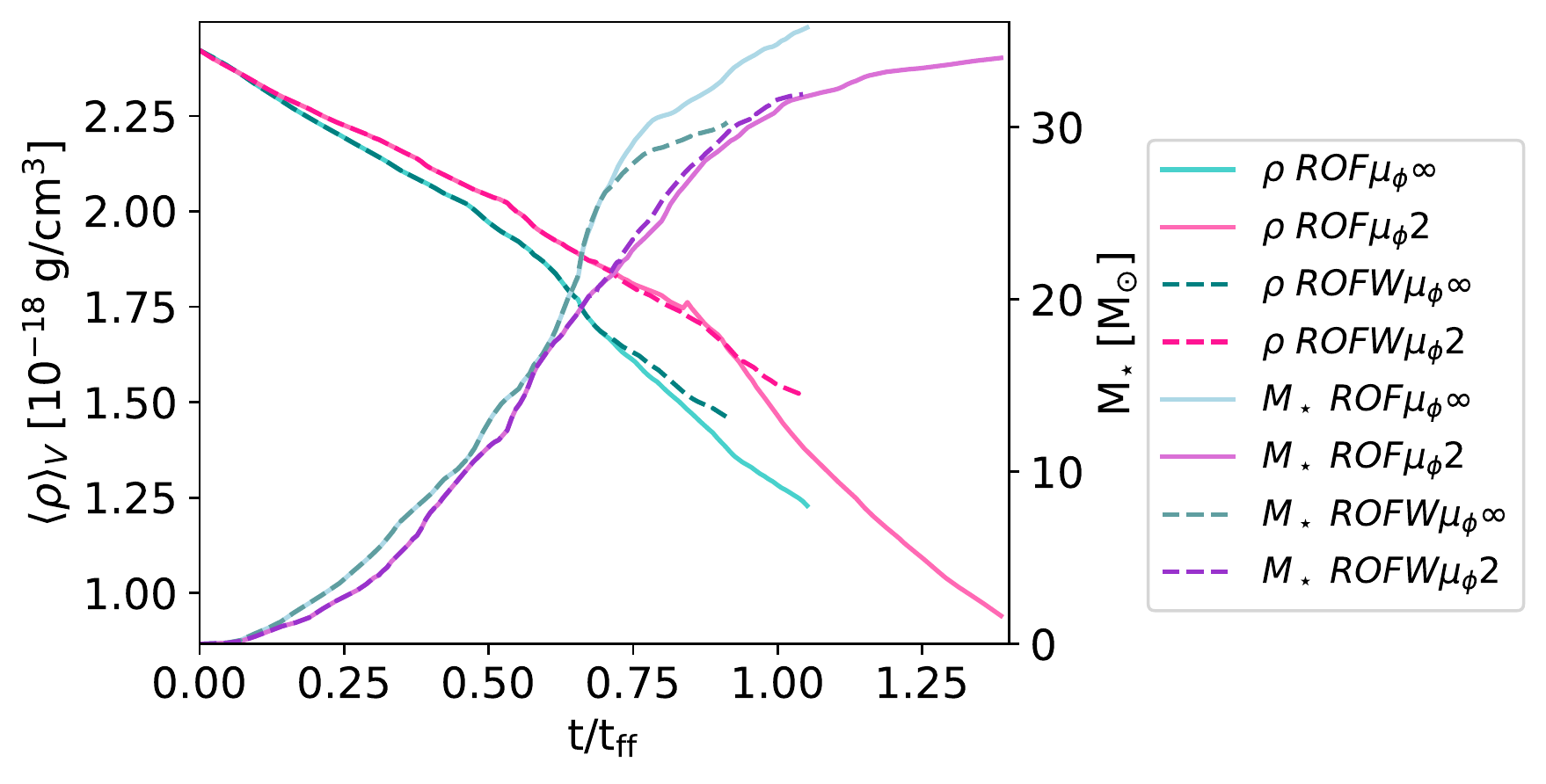}}
\caption{
\label{fig:aveRho}
Volume-weighted average density within a sphere of 0.1 pc (i.e., the initial core radius) as a function of simulation time (left panel) 
for runs  \ofnob\ (light teal solid line), \wofnob\ (teal dashed line), \ofb\ (light pink solid line), and \wofb\ (dark pink dashed line). The left panel also shows the evolution of the primary stellar mass (left $y-$axis) for runs  \ofnob\ (light blue solid line), \wofnob\ (dark blue dashed line), \ofb\ (light purple solid line), and \wofb\ (purple dashed line).
}
\end{figure}

The decrease (increase) in the accretion rate for run \wofnob\ (\wofb) can be quantified by comparing the evolution of the volume-weighted mean density to the stellar mass growth as shown in Figure~\ref{fig:aveRho}. This figure shows the volume-weighted core density, including contributions from both the outflow and wind material (note that these quantities are low compared to the non-accreted core material), within a sphere of 0.1 pc (i.e., the initial core radius) as a function of simulation time for runs \ofnob, \ofb, \wofb, and \wofnob. The evolution of the massive star's mass is over-plotted. As the massive star grows in mass the average density decreases for all runs. This decrease is greater for runs without magnetic fields because the massive star's have a faster growth rate. The average density slightly increases for runs \wofnob\ and \wofb\ as compared to runs \ofnob\ and \ofb\ when the mass growth for the massive stars plateau because wind feedback becomes important. This slight increase occurs earlier for run \wofnob\ as compared to run \wofb\ because the accretion rate drops earlier due to the formation and expansion of the wind-driven bubbles.

The top left middle panel, bottom left middle panel, and bottom right middle panel of Figure~\ref{fig:star_props} shows the primary star's accretion luminosity given by Equation~\ref{eqn:Lacc}, internal (stellar) luminosity, and the ratio of these two quantities as a function of simulation time. When the star is less than several $M_{\rm \odot}$ the accretion luminosity is larger than the stellar luminosity, but eventually the stellar luminosity dominates. The top right middle panel of Figure~\ref{fig:star_props} shows the radial evolution of the massive protostars. The difference in the early radial evolution for these stars is due to their different accretion histories since the accretion of material alters the stellar entropy distribution \citep{Hosokawa2009a}. In agreement, observations of massive protostars by \citet{Ginsburg2017a} find that massive stars are likely bloated as they accrete most of their mass. Until winds become important, the accretion rate is higher for runs \ofnob\ and \wofnob\ and therefore it is more bloated at early times as compared to runs \ofb\ and \wofb. Likewise, when the winds are launched (at around $\sim 20 \; M_{\rm \odot}$) the massive stars are already contracting to the ZAMS so the reduced (increased) accretion flow in run \wofnob\ (\wofb) has a negligible effect on the radial evolution of the massive (proto)stars. The top right-hand and bottom right-hand panels show the accretion timescales, $t_{acc} = M_{\rm \star}/\dot{M}_{\rm \star}$,  and Kelvin-Helmholtz timescales, $t_{\rm KH}=GM^2_{\rm \star}/R_{\rm \star}L_{\rm \star}$, respectively. Once the star is sufficiently massive we find that $t_{\rm KH}< t_{\rm acc}$ owing to the high stellar luminosity and contraction to the main-sequence.

\subsection{Massive Star Wind Properties}
\label{sec:windprops}
\begin{figure*}
\centerline{\includegraphics[trim=0.2cm 0.2cm 0.2cm 0.2cm,clip,width=0.9\textwidth]{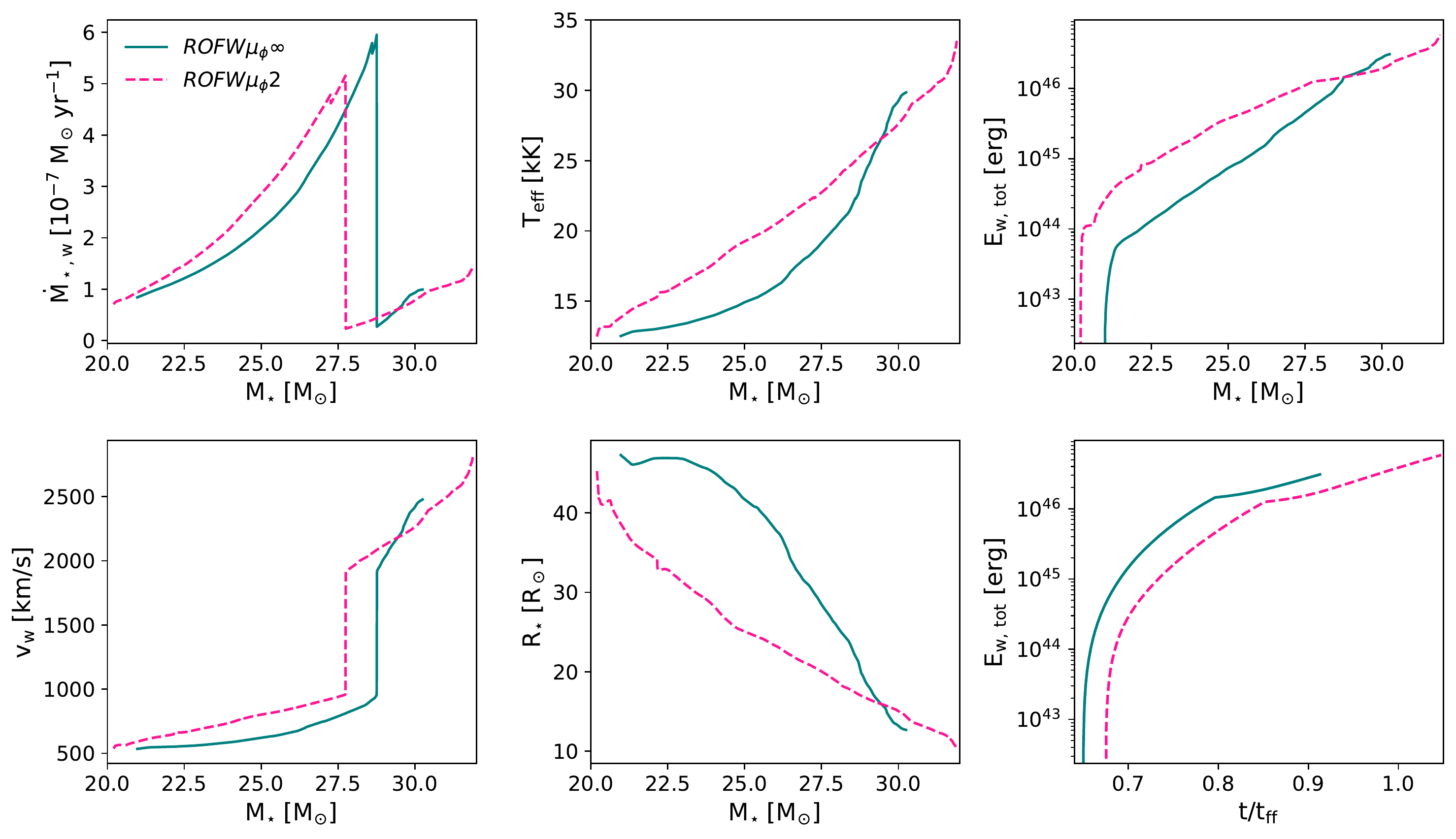}}
\caption{
\label{fig:wind_props_mstar}
Stellar wind properties as a function of primary stellar mass for runs \wofnob\ (solid teal lines) and \wofb\ (dashed pink lines). The top-left panel and bottom-left panel show the wind mass-loss rates and wind velocities following the formulae presented in Section~\ref{sec:winds} when the star has $T_{\rm eff} \ge 12.5$~kK following \citet{Vink2001a}. The top-center and bottom-center panels show the primary star's effective temperature and radius, respectively. The top-right and bottom-right panels show the total integrated wind kinetic energy as a function of stellar mass and simulation time, respectively.
}
\end{figure*}

The wind mass-loss rates following Equations~\ref{eqn:mdot1} and \ref{eqn:mdot2} (top left panel), wind velocities (bottom left panel), stellar effective temperature (center top panel), stellar radius (center bottom panel), and integrated wind kinetic energy ($E_{\rm w, \, tot} = \int \frac{1}{2} \dot{M}_{\rm w} v_{\rm w}^2 \, dt$, top right panel) are shown in Figure~\ref{fig:wind_props_mstar} as a function of stellar mass for the primary star, considering only stellar masses at which the effective temperature is  $\gtrsim 12.5$~kK (i.e., hot enough so that winds are launched) for runs \wofnob\ and \wofb. The bottom right panel also shows the integrated wind kinetic energy ($E_{\rm w, \, tot}$) as a function of simulation time for comparison. 

Initially, the wind mass-loss rates and wind velocities are on the cool-side of the bi-stability jump, as described in Section~\ref{sec:winds} and \citet{Vink2001a}, and are of order $\dot{M}_{\star,w} \sim 10^{-7} \; \rm M_{\rm \odot} \, yr^{-1}$. When winds are initially launched the wind velocities are relatively low (i.e., $v_{\rm w} \sim 500 \; \rm{km/s}$) due to the stars' bloated radii, however the wind velocities steadily increase to $\sim10^3 \; \rm{km/s}$ as the protostars contract to the ZAMS. These quantities steadily increase until the stellar mass reaches $\sim 29$~$\rm M_{\rm \odot}$ for run \wofnob\ and $\sim 28$~$\rm M_{\rm \odot}$ for run \wofb\ because the stars transition to the hot side of the bi-stability jump, thereby causing the mass-loss rates to drop by a factor of $\sim5$ and the wind velocities to increase by a factor of $\sim2$ when this transition occurs. After this transition the mass-loss rates and wind velocities steadily increase as the stars continue to contract to the ZAMS.

The primary star in run \wofb\ transitions to the hot side of the bi-stability jump at a lower stellar mass, as compared to run \wofnob, because the overall stellar accretion rate is lower throughout the simulation time, thereby allowing the star more time to contract to the ZAMS. Eventually, since the growth rate of the massive star in run \wofnob\ is diminished due to wind feedback, the wind mass-loss rate and wind velocity for run \wofnob\ becomes larger than that of run \wofb\ because the massive star, at the same stellar mass, has a smaller radius and therefore a higher effective temperature and luminosity. Hence, we find that the wind properties are highly dependent on the protostellar evolution of the massive star and they evolve with time and stellar mass. Future studies that include wind feedback in the context of massive star formation should account for these effects rather than treat the mass-loss rates and wind velocities as a constant value, which has been neglected in previous theoretical studies modeling wind bubble formation \citep[e.g., ][]{Geen2020a, Geen2021a}.

As described in Section~\ref{sec:bubble}, sustained and expanding wind bubbles in run \wofnob\ form earlier than those in run \wofb. Comparison of the total injected wind energy (lower right panel) as a function of simulation time versus this quantity as a function of stellar mass (top right panel) show that winds from the massive protostar are initially launched for run \wofnob\ earlier than that of \wofb. This earlier onset of wind feedback results in a greater amount of wind energy injected as a function of time even though the total injected wind energy for run \wofb\ is larger as a function of stellar mass due to the faster contraction for the massive protostar. However, since wind feedback reduces accretion onto the massive protostar in run \wofnob\ earlier than run \wofb\ the massive protostar begins to contract more quickly than the protostar in run \wofb\ at $\sim29$~\msun.

\begin{figure*}
\centerline{\includegraphics[trim=0.25cm 0.25cm 0.25cm 0.25cm,clip,width=1\textwidth]{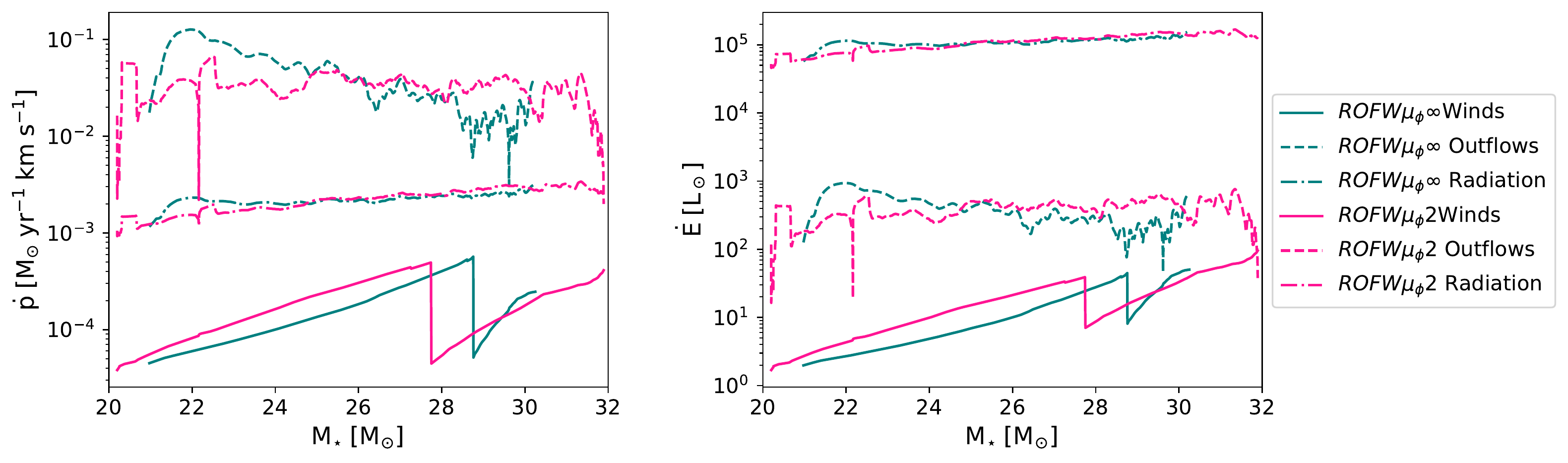}}
\caption{
\label{fig:fb_props_mstar}
Stellar feedback momentum (left-panel) and energy (right-panel) deposition rates as a function of primary stellar mass for stellar winds (solid lines), direct radiation including contributions from both the internal (stellar) and accretion luminosities (dot-dashed lines), and collimated outflows (dashed lines) for runs \wofnob\ (teal lines) and \wofb\ (pink lines), respectively. We show these properties when the star becomes hot enough ($T_{\rm eff} \ge 12.5$~kK) to launch isotropic winds following the mass-loss rates formulae from \citet{Vink2001a}.
}
\end{figure*}

\subsection{Stellar Feedback Comparison}
Figure~\ref{fig:fb_props_mstar} compares the rate of momentum (left panel) and energy (right panel) injection from radiation, collimated outflows, and stellar winds for runs \wofnob\ and \wofb\ as a function of primary stellar mass. Here, the rate of momentum deposited is $\dot{p}_{\rm rad} =(L_{\rm \star} + L_{\rm acc})/c$, $\dot{p}_{o} = \dot{M}_o v_{o}$, and $\dot{p}_{w} = \dot{M}_w v_{w}$  for the direct (stellar+accretion) radiation, outflows, and winds, respectively. Likewise, the rate of energy deposited by these feedback processes are $\dot{E}_{\rm rad} = L_{\rm \star} + L_{\rm acc}$, $\dot{E}_{o} = \frac{1}{2}\dot{M}_o v^2_{o}$, and $\dot{E}_{w} = \frac{1}{2}\dot{M}_w v^2_{w}$. We only consider the kinetic energy from winds and outflows because they dominate over the thermal energy injected by these feedback processes (i.e., $\dot{E}_{\rm Th, \,w}\approx 10^{-3}\dot{E}_{\rm K,\, w}$ and $\dot{E}_{\rm Th, \,o}\approx10^{-2}\dot{E}_{\rm K,\, o}$ for winds and outflows when the primary star is $\gtrsim 20 M_{\rm \odot}$, respectively). These quantities are calculated with the self-consistent primary protostar's properties shown in Figures~\ref{fig:star_props} and \ref{fig:wind_props_mstar}. 

We find that the momentum injected by outflows dominates over the momentum injected by radiation and winds whereas the rate of energy deposited by radiation dominates over that of outflows and winds. Likewise, the momentum and energy injected by outflows is much larger than those injected by stellar winds due to the much higher outflow mass-loss rates even though the wind speeds are higher. Regardless, the size scales of these processes differ. The absorption of the direct radiation field depends on the optical depth of the material it interacts with and the hot-gas produced by the shock-heating of stellar winds, which is effectively transparent to the stellar radiation field because it reaches temperatures much higher than the dust-sublimation temperature ($T_{\rm sub} \approx 1500$~K). As shown in Figures~\ref{fig:densitySlices_NoBzoom} and \ref{fig:densitySlices_Bzoom}, we find that wind feedback drives low-density bubbles near the star before radiation pressure can drive radiation-pressure driven bubbles. Therefore, once wind feedback produces these adiabatic wind bubbles most of the stellar radiation will be absorbed in the dense bubble shells. 

Collimated outflows are injected near the star, similarly to winds, but only over a small covering angle. The momentum in these outflows entrain material that is eventually ejected from the core. We note that although the outflows contain more kinetic energy than winds, they are not efficiently thermalized like stellar winds because of the much lower outflow velocities and therefore the majority of the kinetic energy in outflows is likely lost via radiative cooling \citep{Koo1992a, Rosen2020a, Rosen2021a}. In contrast, the effect of stellar wind feedback is more localized because the injected fast flowing wind material is thermalized close to the star and the resulting hot gas expands adiabatically near the massive star. Therefore, as shown by the density distribution near the massive star in Figures~\ref{fig:densitySlices_NoBzoom} and \ref{fig:densitySlices_Bzoom}, wind feedback affects the gas near the star and therefore is more effective at reducing the accretion flow onto the massive star at late times for runs \wofnob\ and \wofb. Hence, although the energetics of wind feedback is sub-dominant, we find that winds may be more effective at halting accretion onto massive stars because they act more localized as compared to radiation and outflows.

\subsection{Entrained Wind Material and Outflows}
\label{sec:entrained}
As noted in Section~\ref{sec:feedback} we add passively advected scalars (i.e., tracer fields) to the wind and outflow material that is injected, which is used to measure $\rho_{\rm w}$ and $\rho_{\rm OF}$ precisely for each cell.  We define entrained wind and entrained outflow material as consisting of all cells whose mass contains at least 0.005\% and 5\% of the launched wind and outflow material, respectively (i.e., cells where $f_{\rm t,w} = \rho_{\rm w}/\rho \ge 5 \times 10^{-5}$ and  $f_{\rm t,OF} = \rho_{\rm w}/\rho \ge 5 \times 10^{-2}$). We choose a much lower tracer fraction ($f_{\rm t,w}$) for winds, as compared to outflows, because the wind mass-loss rates are several orders of magnitude lower than the launched outflow mass-loss rates. 

\begin{figure*}
\centerline{\includegraphics[trim=0.4cm 9.1cm 0.2cm 0.2cm,clip,width=1\textwidth]{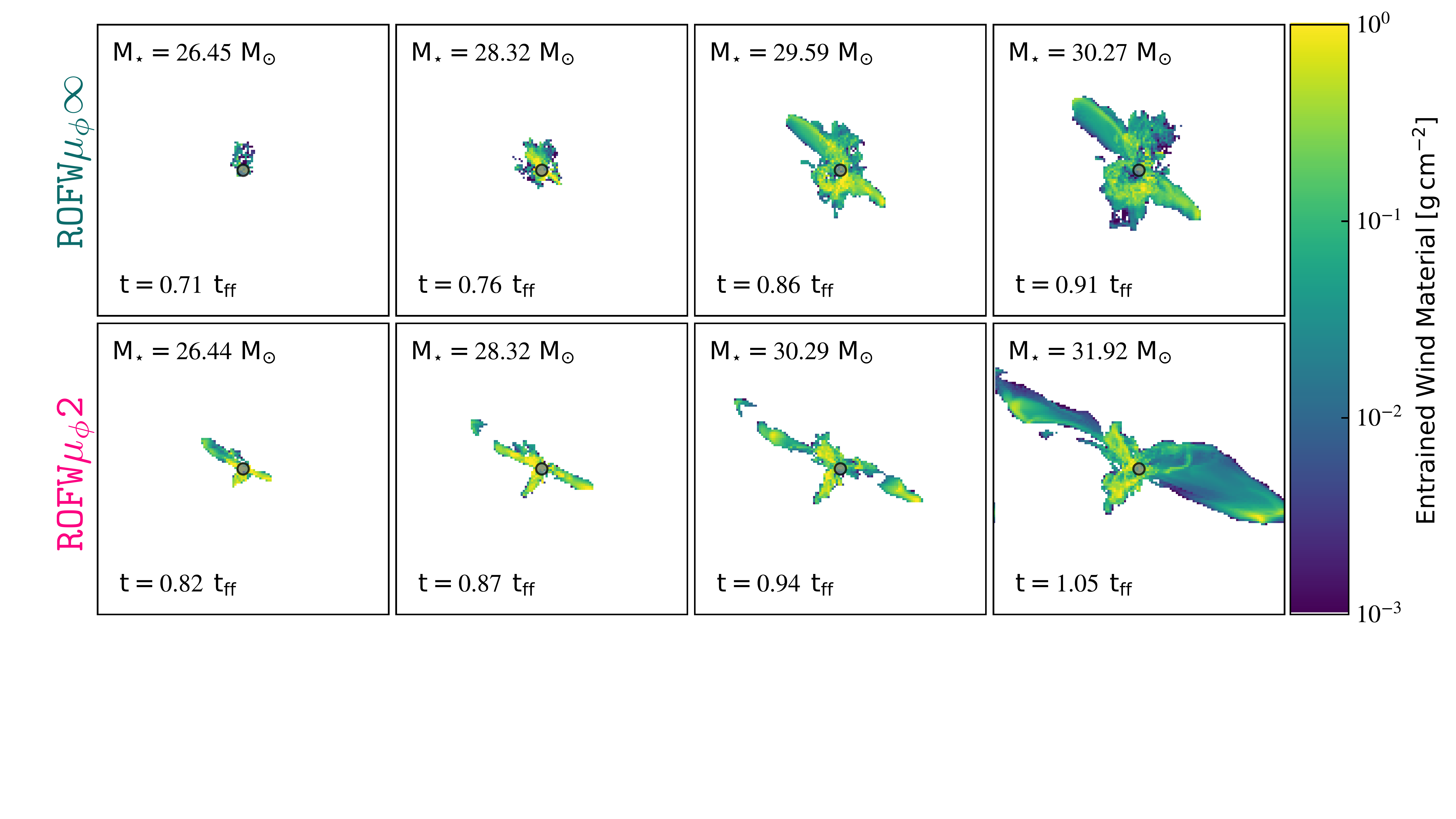}}
\caption{
\label{fig:WindEntrained}
Projections of the density of the entrained wind material along the $yz$-plane that are moving away from the primary star ($v_{\rm rad}>0$) for runs \wofnob\ (top row) and \wofb\ (bottom row). Each panel is (0.4~pc)$^2$
}
\end{figure*}

\begin{figure*}
\centerline{\includegraphics[trim=0.4cm 9.1cm 0.2cm 0.2cm,clip,width=1\textwidth]{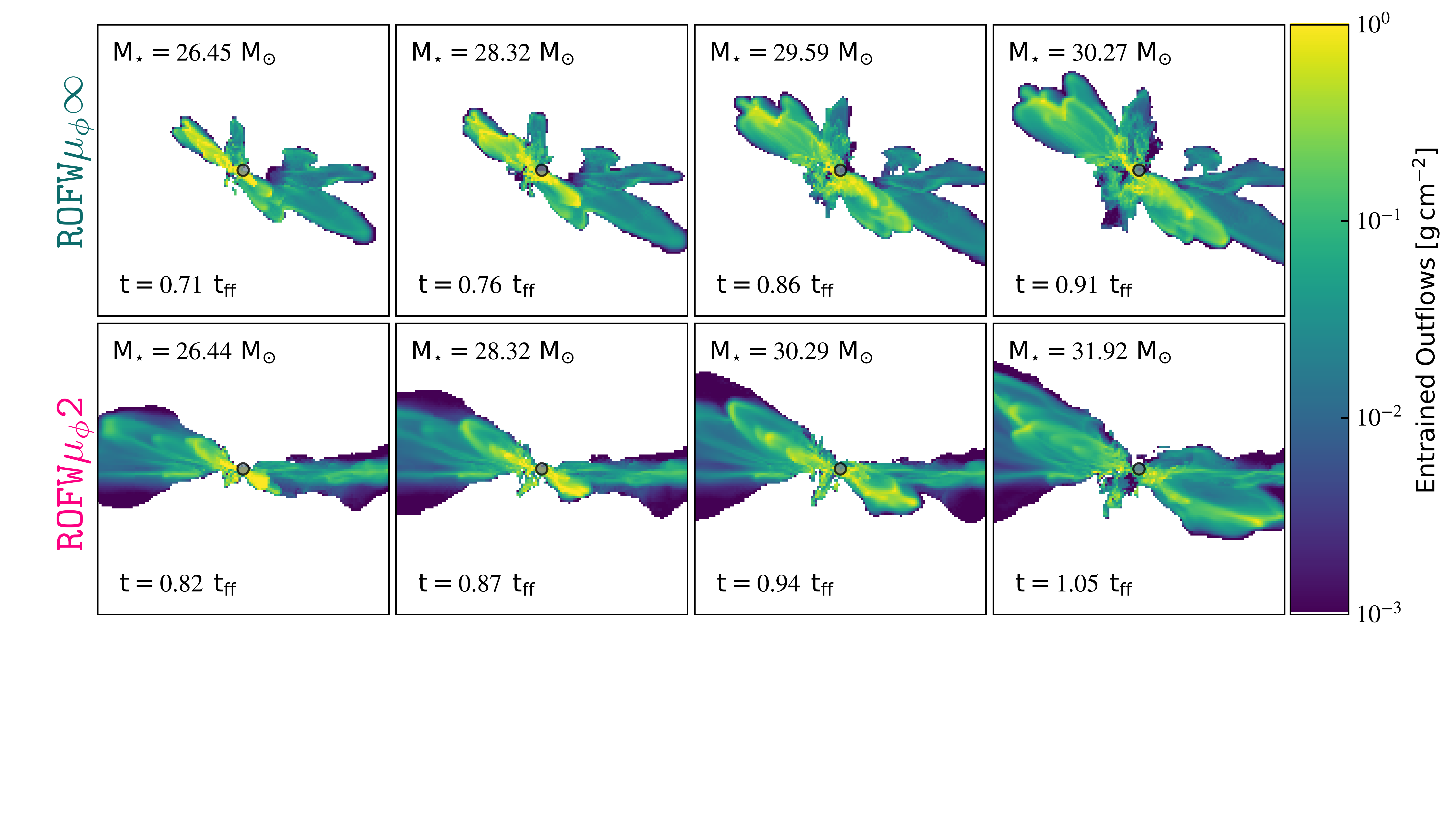}}
\caption{
\label{fig:OutflowsEntrained}
Projections of the density of the entrained outflow material along the $yz$-plane that are moving away from the primary star ($v_{\rm rad}>0$) for runs \wofnob\ (top row) and \wofb\ (bottom row). Each panel is (0.4~pc)$^2$
}
\end{figure*}

Figure~\ref{fig:WindEntrained} shows the evolution of the surface density of the entrained wind material for runs \wofnob\ (top row) and \wofb\ (bottom row). We note that we do not impose any temperature cuts when integrating over cells that contain wind material and therefore we are highlighting the advection of the wind material rather than only the hot energy-driven wind bubbles discussed in Section~\ref{sec:bubble}. These panels demonstrate that the core and outflow material that is entrained by winds has a roughly bipolar morphology that expands  as the primary star grows in mass. We compare these with the evolution of the surface density of the entrained molecular outflows, as shown in Figure~\ref{fig:OutflowsEntrained}, which also has a bipolar morphology due to the collimated structure of the outflows that are launched by the primary star. 

Comparison of these two figures show that the wind material primarily expands along the low-density gas that is carved out by outflows even though the stellar winds are launched isotropically. Additionally, comparison of the last column in Figure~\ref{fig:WindEntrained} shows that the entrained wind material for run \wofb\ encompasses a larger volume than run \wofnob. This suggests that the wind material undergoes the magnetic levitation effect similar to the entrained outflows, as shown in Figure~\ref{fig:OutflowsEntrained}, which was first described in  \citet{Rosen2020b}.

\begin{figure*}
\centerline{\includegraphics[trim=0.4cm 9.1cm 0.2cm 0.2cm,clip,width=1\textwidth]{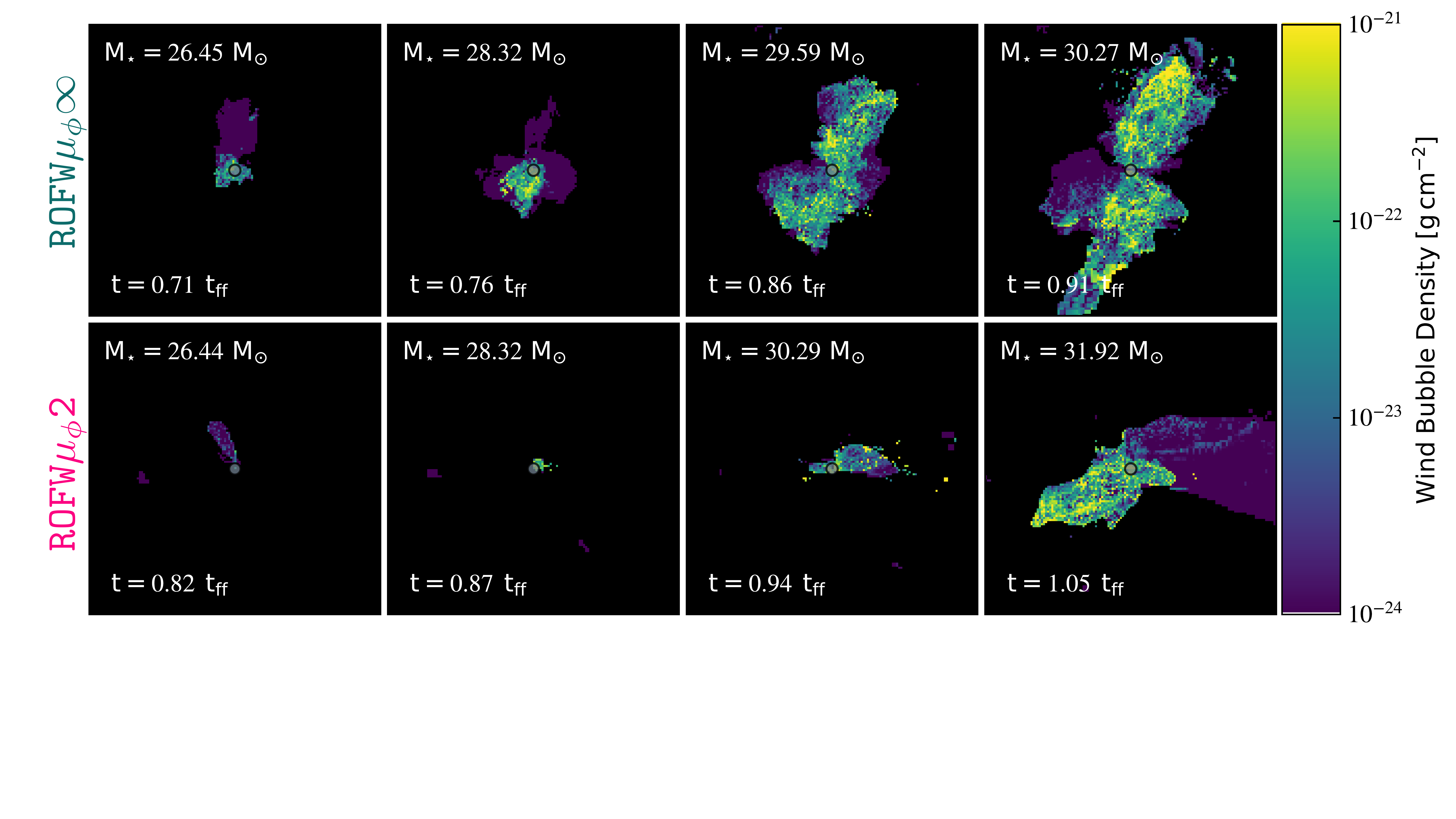}}
\caption{
\label{fig:windBubbles}
Projections of the density of the wind bubbles along the $yz$-plane that are moving away from the primary star ($v_{\rm rad}>0$) for runs \wofnob\ (top row) and \wofb\ (bottom row) for gas with $T\ge 10^4$~K. Each panel is (0.15~pc)$^2$.
}
\end{figure*}

Figure~\ref{fig:WindEntrained} considered all gas that is entrained by the wind material injected by the massive star. However, as described in Section~\ref{sec:bubble}, we found that the thermalization of stellar winds leads to hot gas that adiabatically expands leading to hot wind bubbles that eventually reduces the accretion rate onto the massive star. In order to determine if these hot wind bubbles preferentially expand along the same direction as the low-density gas carved out by outflows, we show projections of the mass-weighted density of the hot and warm gas ($T \ge 10^4$~K) generated by stellar wind feedback in Figure~\ref{fig:windBubbles} for runs \wofnob\ (top row) and \wofb\ (bottom row). We chose a temperature cutoff of $T \ge 10^4$~K because, even though the shock-heated gas produced by wind feedback has $T \ge 10^6$~K, this gas will eventually mix with the surrounding cool turbulent gas as it expands that then rapidly cools via turbulent mixing and conduction \citep{Rosen2014a, Lancaster2021b}. These projections show that the edges of the bubbles are highly turbulent, thereby confirming that turbulent mixing occurs at the hot-cold interface of the bubble shells.

We find that the wind driven bubbles are roughly bipolar but do not lie along the same orientation as the majority of the entrained wind and outflow material. Likewise, the wind bubble lobes for the last snapshot of run \wofnob\ show the pinched morphology near the star due to the shielding of the accretion disk. This is less apparent for run \wofb\ because a substantial accretion disk does not form due to magnetic braking of the infalling material. Regardless, the wind bubble in run \wofb\ is not spherical and instead is elongated. Comparison of Figures~\ref{fig:OutflowsEntrained} and \ref{fig:windBubbles} show that these lobes do not lie along the same direction as the entrained outflows. Hence, we find that wind material more easily expands along regions where outflow feedback carves out low-density gas but that the hot wind bubbles produced by the shock-heating of stellar winds do not primarily expand along the same direction and are instead more localized near the star. Additionally, the hot gas produced by wind feedback preferentially expands along directions that are unimpeded by the dense circumstellar material near the massive star.

\begin{figure}
\centerline{\includegraphics[trim=0.2cm 0.2cm 0.2cm 0.2cm,clip,width=1\columnwidth]{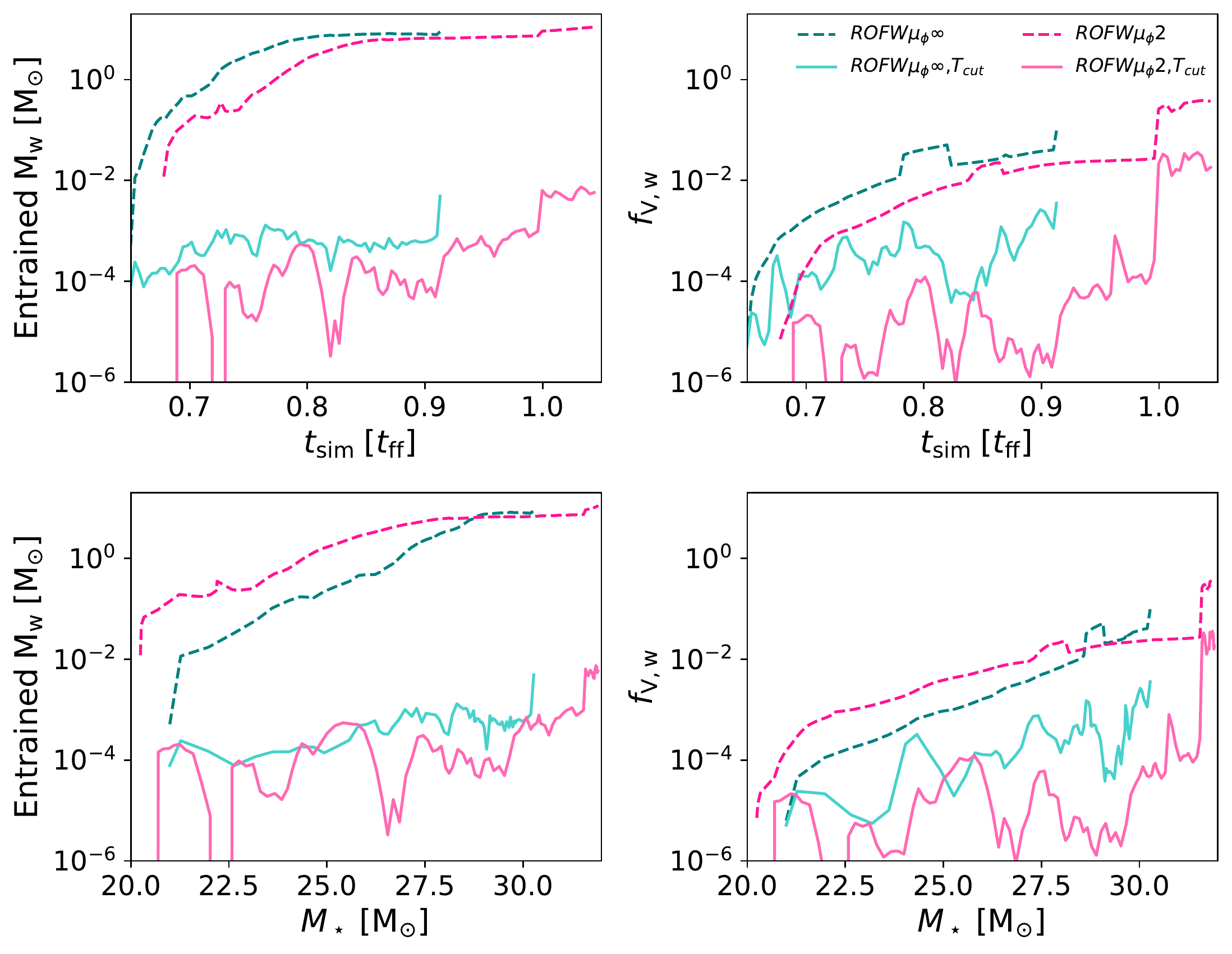}}
\caption{
\label{fig:windEnt}
Entrained wind material (left panels) and wind-material volume filling factor (right panels) as a function of simulation time (top row) and primary stellar mass (bottom row). The lines show where these quantities contain $5\times10^{-3}\%$ of wind material. The dark teal and dark pink dashed lines consider all gas for runs \wofnob\ and \wofb, respectively. The light teal and light pink solid lines only consider gas with $T\ge10^4$~K for runs \wofnob\ and \wofb, respectively.
}
\end{figure}

We quantify these effects in Figure~\ref{fig:windEnt}, which shows the entrained wind material (left column) and its volume filling fraction (right column) as a function of simulation time (top panels) and primary stellar mass (bottom panels) for runs \wofnob\ and \wofb. We calculate the volume filling fraction by summing over all cells whose mass contains launched wind material normalized to the initial core volume: $f_{\rm w,V} = \sum_i dV_{\rm{w},i}/V_{\rm core, \, init}$ where $V_{\rm core, \, init} = \frac{4}{3}\pi R^3_c$. The dot-dashed lines consider all cells with $f_{\rm t,w} = \rho_{\rm w}/\rho \ge 5 \times 10^{-5}$ whereas the solid lines only consider cells with $T\ge10^4$~K (i.e., corresponding to the wind bubbles shown in Figure~\ref{fig:windBubbles}). This figure demonstrates that entrained wind material consists of primarily cold ($T<10^4$~K) gas that encompasses a larger volume than the hot adiabatic wind bubbles. The hot ($T\gtrsim10^4$~K) wind bubbles encompass a very small volume-filling factor in comparison because they are initially crushed by the surrounding infalling gas, but once the massive protostar reaches $\sim30$~\msun\ the bubbles exhibit sustained growth and increases in size. At this point for run \wofb\ the rapid bubble expansion transitions to a more spherical morphology  because this simulation doesn't form a noticeable accretion disk.

\begin{figure*}
\centerline{\includegraphics[trim=0.2cm 0.35cm 0.2cm 0.2cm,clip,width=1\textwidth]{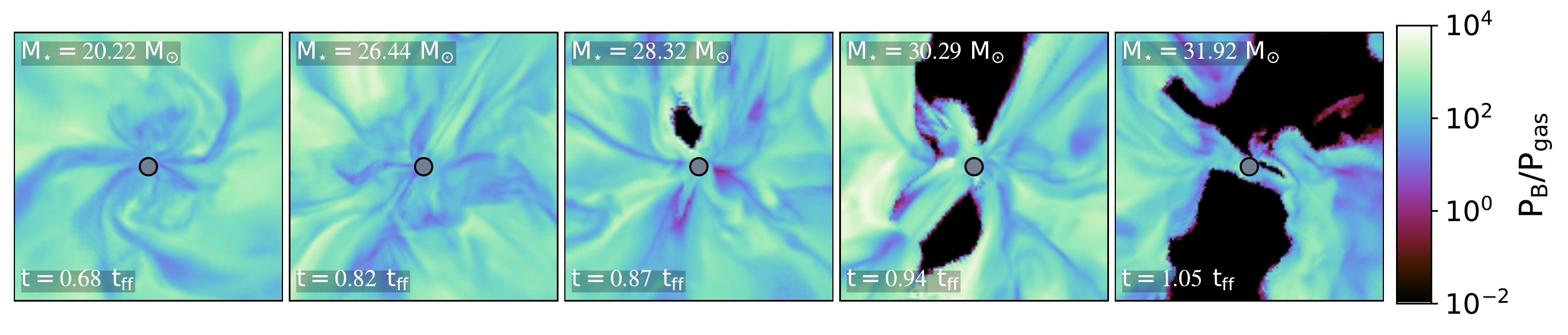}}
\caption{
\label{fig:pressure}
Thin density-weighted projections of the ratio of the magnetic pressure to thermal gas pressure ($P_{\rm B}/P_{\rm gas}$) for run \wofb. Each panel is (5000~au)$^2$ in area and the projection is taken over a depth of 500~au with the massive star, denoted by the gray circle, at the center. Each projection is oriented such that the mass-weighted angular momentum axis of the gas within a radius of 500~au from the massive star points out of the page.
}
\end{figure*}

\subsection{Role of Magnetic Tension in Combating Wind Feedback}
\label{sec:bfields}
In this work we found that magnetic fields have a direct influence on the importance of wind feedback in massive star formation.  As noted before, analytic theory and numerical simulations have demonstrated that the fast wind material becomes thermalized and the resulting  gas expands adiabatically due to its high temperature \citep{Weaver1977a, Koo1992a, Rosen2021a}. However, this effect is diminished at early times for run \wofb\ as compared to run \wofnob, which does not include magnetic fields.

Figure~\ref{fig:pressure} shows thin density-weighted projections of the ratio of the magnetic pressure ($P_{\rm B}$) to thermal gas pressure ($P_{\rm gas}$) for run \wofb\ as a function of simulation time. These panels show that throughout most of the simulation (once winds are launched) the material near the massive star has $P_{\rm B} \gg P_{\rm gas}$. Eventually regions near the massive star (i.e., when we have wind-driven bubbles that expand away from the star) have thermal gas pressure much larger than the magnetic pressure. Hence, until the winds become significant the magnetic tension associated with the magnetized material near the star reduces the thermalization of the stellar wind material and also confines the expansion of the resulting hot gas. Therefore, we find that confinement due to magnetic tension likely reduces the production and expansion of wind-driven bubbles that form via wind feedback, which are more apparent in run \wofnob\ since the wind-driven bubbles are launched at a lower stellar mass.  

\begin{figure}
\centerline{\includegraphics[trim=0.2cm 0.35cm 0.2cm 0.2cm,clip,width=1\columnwidth]{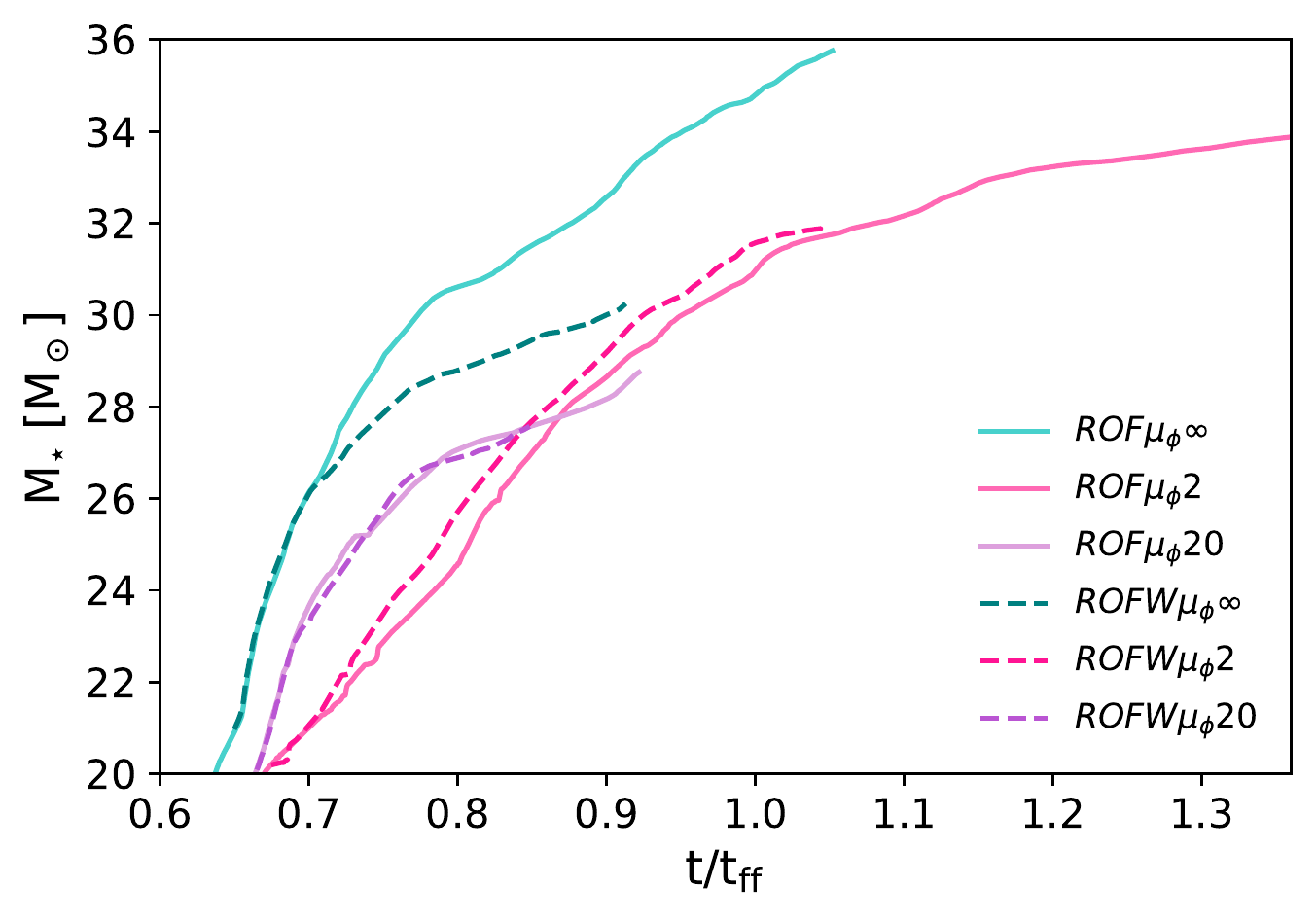}}
\caption{
\label{fig:mstarAll}
Primary protostar mass as a function of simulation time for runs \ofnob,\wofnob, \ofb, \wofb, \ofsb, and \wofsb.
}
\end{figure}

In order to determine how magnetic fields influence wind feedback we compare two identical simulations to runs \ofb\ and \wofb\ but include a magnetic field strength that is a factor of 10 lower yielding an initial $\mu_{\phi} = 20$ (runs \ofsb\ and \wofsb), which acts as an intermediate case to the magnetized and unmagnetized cores described throughout this work. Figure~\ref{fig:mstarAll} shows the growth rate for the massive star that forms in all of the simulations as a function of simulation time. As described in Section~\ref{sec:starprops} we found that wind feedback reduces the mass growth of the primary star when the core is not magnetized, but the mass growth is slightly enhanced when it is. Comparison of runs \ofsb\ and \wofsb\ with the other simulations shows that a weaker magnetic field reduces the mass growth compared to the non-magnetic case but increases it compared to the magnetic case. When winds become important for run \wofsb\ we find that the mass growth rate is very similar to the mass growth of the massive star in run \ofsb\ suggesting that if the core is weakly magnetized then winds do not enhance or decrease the growth rate of the massive star at least for the time simulated here.

\section{Discussion}
\label{sec:disc}
The purpose of this work is to understand how feedback from isotropic radiatively-driven stellar winds, in concert with feedback from collimated outflows and radiation pressure, affect the formation of massive stars that form from the gravitational collapse of unmagnetized and magnetized massive prestellar cores. Most notably, we find that wind feedback significantly reduces accretion onto massive stars that form from the collapse of unmagnetized prestellar cores. However, for collapsing magnetized cores we find that accretion onto massive stars is slightly enhanced before it begins to taper off once winds are powerful enough to launch adiabatic wind bubbles since magnetic tension in the surrounding material inhibits their growth at early times. Additionally, we find that the wind-blown bubbles that emanate are not spherical and are instead roughly bipolar because the hot gas is pinched by the dense circumstellar material or disk that forms around the massive star. In this section, we discuss a new phenomenon called the ``wind tunnel effect" that describes this bipolar structure of the wind-driven bubbles in Section~\ref{sec:tunnel}, the implications of studying these structures with \chandra\ and future X-ray telescopes in Section~\ref{sec:obs},  and we address the caveats of our simulations in Section~\ref{sec:caveats}. 


\subsection{The ``Wind Tunnel Effect"}
\label{sec:tunnel}
Previous numerical work that modeled the role of radiation pressure in massive star formation, but neglected wind feedback, demonstrated that the presence of an optically thick accretion disk reduces the effects of radiative acceleration in the radial direction leading to the ``flashlight effect" in which the radiative flux escapes along the polar axis and into the polar cavities, launching radiation-pressure-dominated bubbles above and below the star \citep[e.g.,][]{Yorke2002a, Krumholz2009a, Rosen2016a, Rosen2019a}. The simulations presented here also include isotropic wind and collimated outflow feedback from massive stars, and show a similar effect occurs for the hot shock-heated gas (T$\gtrsim 10^6$~K) produced by stellar wind feedback that undergoes adiabatic expansion (i.e., $P\; dV$ work). The hot gas preferentially expands in directions perpendicular to the accretion disk or dense circumstellar material since the gas density is much lower along these directions. Hence, we find that the thermalized gas produced by wind feedback experiences the greatest expansion along the polar directions of the massive star, which we term the ``wind tunnel effect" in an analogous manner to the ``flashlight effect" due to radiative feedback. 

This ``wind tunnel effect" produces hot gas lobes that are pinched by the dense accretion disk or dense circumstellar material, thereby resulting in a hour glass morphology as shown in Figures~\ref{fig:bubble_noB}, \ref{fig:bubble_B}, and \ref{fig:windBubbles}  for runs \wofnob\ and  \wofb, respectively; rather than a spherical bubble as would be expected for isotropic wind feedback in a uniform medium \citep{Weaver1977a, Koo1992a}. By comparing runs \wofnob\ and  \wofb\ with identical simulations that neglect wind feedback (runs \ofnob\ and  \ofb), we find that the ``wind tunnel effect" occurs before radiation pressure becomes strong enough to launch the radiation-pressure-dominated bubbles. Regardless, the morphology of these two effects are similar and the result is that feedback from radiation and winds, both of which are launched isotropically, lead to asymmetric low-density lobes that expand as the massive star grows in mass. When magnetic fields are included we find that the ``wind tunnel effect" is delayed and the wind-driven lobes are smaller in volume than the non-magnetic case, at a given stellar mass, because the expanding wind bubbles are confined by magnetic tension. This similar effect was found for the radiation-pressure-dominated bubbles presented in \citet{Rosen2020b}.

One interesting comparison to note here is that the structure of the bipolar wind bubbles expected from massive protostars, as demonstrated in this work, are morphologically similar to those observed for evolved massive stars and interacting massive binaries. For example, the nebulae of evolved massive stars such as fast-rotating luminous blue variables (LBVs) typically have a bipolar morphology because equatorial gravity darkening will lead to a higher mass-loss rate and faster wind speed along the poles of the star \citep{Dwarkadas2002a, Lobel2013a, Smith2014a}. Likewise, some Wolf Rayet (WR) stars have also been observed to have a bipolar wind bubble morphology, these structures likely occur if the WR star had undergone a blue supergiant evolutionary phase or experienced binary interactions or mergers during its red supergiant evolutionary phase \citep{Meyer2021a}.


\begin{figure*}
\centerline{\includegraphics[trim=0.3cm 12.5cm 0.2cm 0.2cm,clip,width=1\textwidth]{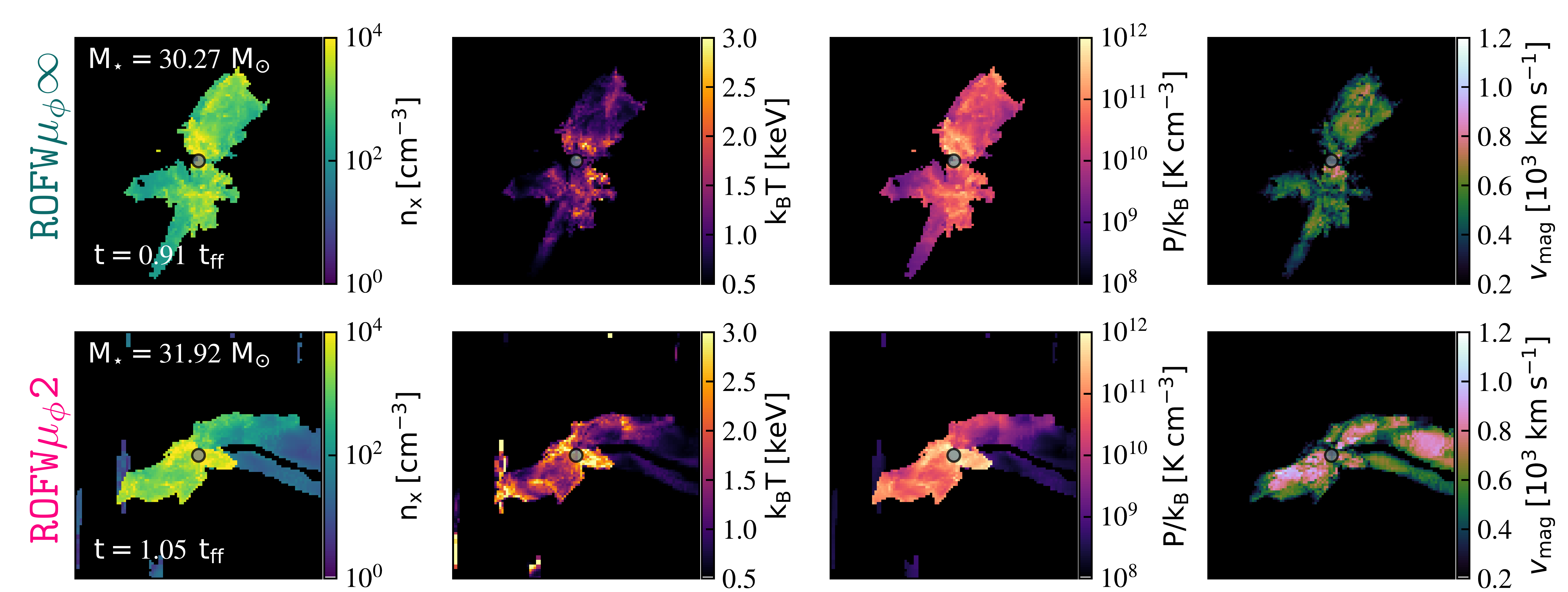}}
\caption{
\label{fig:windBubbleProps}
Projections of the number density (left column), temperature (center left column), pressure (center right column), and velocity magnitude (right column) of the hot gas within the wind bubbles along the $yz$-plane of the final snapshots for runs \wofnob\ (top row) and \wofb\ (bottom row) considering only gas with $T\ge 0.5$~keV ($T\ge 5.8 \times 10^6$~K). Each panel is (0.15~pc)$^2$.
}
\end{figure*}

\subsection{Implications for X-ray Observations}
\label{sec:obs}
Extended X-ray emission has been detected for several ultra-compact and compact \hii\ regions ($R_{\rm \hii} \lesssim 0.5$~pc) in the Milky Way \citep[][]{Takagi2002a, Tsujimoto2006a, Anderson2011a, Skinner2019a, Olivier2021a}. This emission likely traces the hot ($T\gtrsim 10^7$~K) thermal plasma generated by the shock heating of stellar winds launched by individual or multiple massive (proto)stars, but the X-ray emission may also be affiliated with the central sources \citep[i.e., the massive protostars that power the winds and nearby low-mass pre-main sequence stars;][]{Feigelson2005a, Getman2005a}. \chandra\ observations of these compact \hii\ regions found that the emission is dominated by hard X-rays ($\ge 3$~keV) and these observations were unable to resolve the stellar sources from the diffuse plasma, thereby leading to confusion \citep{Olivier2021a}. The soft X-rays ($\lesssim 3$~keV) were likely highly attenuated by the high column densities ($N_{\rm H} \gtrsim 10^{23} \; \rm cm^{-2}$) of the compact \hii\ regions \citep{Takagi2002a, Tsujimoto2006a, Anderson2011a}. In order to shed light on these observations and understand why the detected \chandra\ X-ray emission is dominated by $\gtrsim 3$~keV photons we look at the distribution of the hot gas properties with $k_{\rm B}T_{X} \ge 0.5$~keV ($T_{X} \ge 5.8$~MK) that could be detectable by \chandra. 

Figure~\ref{fig:windBubbleProps} shows the density-weighted projections of the number density ($n_X$; left column), temperature ($k_{\rm B}T_X$; center left column), thermal pressure ($P_X/k_{\rm B}=1.9 n_X T_X$ where the factor of 1.9 assumes that He is doubly ionized and the He mass fraction is 0.25), and velocity magnitude ($v_{\rm mag}$) of the hot gas produced by the shock heating of stellar winds for the final snapshots for runs \wofnob\ (top row) and \wofb\ (bottom row). These panels show that this hot gas occupies a small filling factor ($f$; i.e., the volume of the \hii\ region occupied by the hot shock-heated gas) for the region. \citet{Olivier2021a} assume a filling factor $f=1$ when estimating $P_{X}$ (i.e., the hot gas fully occupies a sphere with radius $R_{\hii}$), however we find that $f\approx 0.02$ for both simulations (assuming $R_{\hii} = 0.075$~pc) suggesting that the hot gas should only occupy a tiny non-spherical fraction of the (assumed) spherical \hii\ regions. 

The volume-weighted average wind bubble properties and intervening core column densities for the final snapshots of runs \wofnob\  and \wofb\ are listed in Table~\ref{tab:xrays}.  We use \chianti\ \citep{Dere1997a} to compute the emissivity, $j_{\nu}(T,Z)$, to obtain the total wind bubble X-ray luminosity $L_{X} = 0.9 n_x^2 j(T_X,Z)$ where $j(T_X,Z)$ is the integrated total emissivity over the ($0.5-7$)~keV X-ray band, $T_{X}$ and $n_X$ are the volume-weighted average quantities listed in Table~\ref{tab:xrays}, and we assume solar metallicity \citep{Rosen2014a}. We compute the flux, $F_{X} = L_{X}/(4\pi D^2)$, by assuming a distance of $D=11.1$~kpc \citep[][i.e., the distance to the massive star forming region W49A that had detected X-ray emission in \citet{Tsujimoto2006a} and \citet{Olivier2021a}]{Zhang2013a}. The X-ray photon count rates for the soft (0.5-3~keV) and hard (3-7~keV) \chandra\ wavelength ranges are then computed with \textit{WebPIMMS}\footnote{\blue{\underline{\href{https://heasarc.gsfc.nasa.gov/cgi-bin/Tools/w3pimms/w3pimms.pl}{https://heasarc.gsfc.nasa.gov/cgi-bin/Tools/w3pimms/w3pimms.pl}}}} for the \chandra\ ACIS-I detector\footnote{The ACIS-I detector was chosen because it is better suited to observe compact sources.}  and we assume the flux is dominated with thermal bremsstrahlung emission with $kT$ equal to volume-averaged values listed in Table~\ref{tab:xrays}. The resulting values are listed in Table~\ref{tab:xrays}. In agreement with the observations mentioned above, we find that the soft X-ray band ($\le 3$~keV) would be undetectable by \chandra\ due to the high attenuating column densities of the protostellar core ($N_{\rm H} \gtrsim 4.5 \times 10^{23} \; \rm cm^{-2}$) and that the hard X-ray band ($\ge 3$~keV) should be more easily detectable. 

These results suggest that studying the impact of wind feedback in the early formation of massive stars, when the stars are heavily embedded with $N_{\rm H} > 10^{23} \; \rm{cm^{-2}}$, is challenging. However, as these \hii\ regions expand and evolve, thereby achieving lower intervening attenuating column densities, the diffuse soft X-ray emission will be less attenuated and should be observable with \chandra. Such emission has been observed for a number of extended \hii\ regions ($R_{\hii}\ge$ few~pc) that host massive star clusters \citep{Lopez2011a, Townsley2011c, Rosen2014a, Lopez2014a}. However, future X-ray telescopes with much greater spatial resolution and sensitivity than \chandra\ may be able to differentiate the point-like stellar sources from the compact diffuse wind-bubble emission from the compact highly-embedded \hii\ regions from massive protostars within the Milky Way.

\begin{table}
	\begin{center}
	\caption{
	\label{tab:xrays}
Wind-driven Bubble Properties
}
	\begin{tabular}{ l  c  c}
	\hline
	 & \wofnob\  & \wofb\ \\
	$M_{\rm \star, f}$ $[M_{\rm \odot}]$ & 30.27 & 31.92 \\
	$n_{X}$ $\rm [cm^{-3}]$\tablenotemark{a} & $1.16 \times 10^3$ & $1.21 \times 10^3$ \\
	$T_{X}$ [keV]\tablenotemark{a} & $1.19$ & $1.59$ \\
	$P_{X}/k$ $[\rm K \; cm^{-3}]$\tablenotemark{a} & $4.48 \times 10^{9}$ & $7.56 \times 10^{9}$ \\
	$v_{\rm mag}$ $\rm [km\; s^{-1}]$\tablenotemark{a} & $513.5$ & $703.0$ \\
	$f\tablenotemark{b}$ & $0.0196$ & $0.0221$ \\
	$N_{H}$ $\rm [cm^{-2}]$\tablenotemark{d} & $1.0 \times 10^{24}$ & $4.5 \times 10^{23}$\\
	$L_{X}$ $[\rm 10^{33} \; erg \; s^{-1}$]\tablenotemark{e} & 6.20 & 9.82 \\
	$F_{X}$ $[\rm 10^{-13} \; erg \;  cm ^{-2} \; s^{-1}]$ & 4.21 & 6.66 \\
	$(0.5-3)$~keV CPS $\rm[s^{-1}]$\tablenotemark{f} & $1.083 \times 10^{-8}$  & $ 1.224\times 10^{-5}$  \\
	$(3-7)$~keV CPS $\rm[s^{-1}]$\tablenotemark{f} & $4.099\times 10^{-5}$ &  $6.808\times 10^{-4}$ \\
	\hline
\end{tabular}
\tablenotetext{a}{Volume-weighted quantity.}
\tablenotetext{b}{Hot gas filling factor within a sphere with radius $R=0.075$~pc centered on the massive star.}
\tablenotetext{c}{Average attenuating column density.}
\tablenotetext{d} {Considering emission only from gas with $kT\ge0.5\, \rm keV$.}
\tablenotetext{e} {Unabsorbed flux assuming a distance of $D=11.1$~kpc ($F_{X} = L_X/ 4 \pi D^2$).}
\tablenotetext{f} {Photon counts per second. Calculated via WebPIMMs. }
\end{center}
\end{table}

\subsection{Caveats}
\label{sec:caveats}
In this work, we simulated the impact of stellar feedback from massive protostars that form from the gravitational collapse of \textit{isolated} massive pre-stellar cores and found that the thermalization of stellar winds (i.e., energy-driven wind feedback) can cause the accretion of material to be quenched when the massive star reaches $\sim30$~\msun. However, observations have demonstrated that massive stars form in dense, highly dynamical GMCs that are undergoing gravitational collapse, thereby driving converging flows that provide large-scale accretion to the  birth sites of massive stars \citep{Williams2018a, Kumar2020a, Rosen2020a, Avison2021a}. 
The associated ram pressure associated with these inflows will likely make feedback less important for massive star formation \citep[e.g.,][]{Kuiper2018a}. \citet{Grudic2022a} showed that massive stars tend to form later than their low-mass counterparts in GMCs and that they likely form in high-density sites where dynamical accretion occurs. Therefore, our results demonstrate that massive stars $\gtrsim30$~\msun\ likely can not form from the collapse of isolated massive cores that are not undergoing dynamical accretion from its external environment. Hence, future numerical work studying massive star formation should follow how mass is accumulated from the large (GMC) to small scales (birth sites of massive stars). Studies like these can be done with the new STARFORGE project, which is capable of resolving individual star formation within GMCs and includes all of the relevant stellar feedback processes, which is now studying this in more detail \citep{Grudic2021a, Grudic2022a, Guszejnov2022a}.

In the simulations presented here, which neglected non-ideal MHD effects, we find that magnetic fields have a direct influence on the importance of wind feedback in massive star formation. Since these simulations are in the ideal limit, they assume that the magnetic field lines are well-coupled to the gas and therefore the magnetic field strength is amplified as the core contracts. This leads to a greater magnetic field strength near the massive protostar, where the density is high, thereby resulting in a larger magnetic tension that can inhibit or delay the growth of the adiabatic wind bubbles \citep[e.g.,][]{Rosen2020b}. Non-ideal effects, such as ambipolar diffusion and Ohmic dissipation, likely weaken the magnetic field strength near the massive star \citep{Kolligan2018a, Zhao2020a}, thereby making magnetic tension less important as the system evolves and the star grows in mass. Hence, future work that explores the importance of wind feedback in massive star formation should include these effects. We note that ideal MHD effects are likely a good approximation for the wind bubbles simulated here since the thermalized gas is fully ionized. However, non-ideal MHD effects may weaken the magnetic field near the accreting protostar before winds are launched, thereby reducing the magnetic tension that is responsible for suppressing the growth of the adiabatic wind bubbles that eventually inhibit accretion.

Additionally, we note that the stellar wind modeling used in the simulations presented here assume that winds are initially launched once the protostar reaches an effective temperature of $\sim12.5$~kK following the prescription by \citet{Vink2001a}, which modeled the wind properties of non-accreting massive stars based on their stellar properties and interpolated their model to lower temperatures. Hence, it still remains uncertain when massive (proto)stars begin to experience mass-loss due to radiation pressure in their atmospheres. However, \citet{Vink2018a} modeled the stellar wind properties for very massive, bloated stars with $T_{\rm eff}=15$~kK and found that radiation pressure was able to launch slow winds ($\sim$few $\times 10^2$~km/s) with high mass-loss rates due to their bloated radii and high luminosities. Therefore, we expect stellar winds should initially be launched during the accretion phase for massive protostars due to their high luminosities ($\sim$$10^5-10^6~L_{\rm \odot}$). Regardless, future wind modeling should study how winds can be launched when the massive protostar is actively accreting and contracting to the main sequence. 

Another important detail about radiatively-driven stellar winds that is neglected in these simulations are the small-scale density inhomogeneities (i.e., clumping) within the stellar wind that are a result of the strong, intrinsic instability of line-driving \citep[i.e., the line-deshadowing instability;][]{Sundqvist2018a, Smith2014a, Brands2022a}. This feature of radiatively-driven stellar winds is still poorly understood, therefore we were not able to properly include this effect in our sub-grid wind model. We note that future studies should consider to include wind clumping in their mass-loss prescriptions because asymmetric mass-loss will likely introduce additional asymmetries in the wind bubbles that form around massive protostars.

Finally, our sub-grid wind launching model only included contributions from the stellar surface. \citet{Kee2018b, Kee2019a} found that UV radiation from main-sequence-like massive stars can also ablate the accretion disk leading to an extended supersonic disk wind near the star, thereby enhancing the overall mass-loss rate of the star-disk system. For main-sequence massive stars, they found that the disk ablation rate scales as $\dot{M}_{\rm abl} \sim 6.5 \dot{M}_{\rm \star, \, w}$ and therefore UV-driven disk ablation should reduce the final stellar masses by 15\%. Our results suggest disk ablation may occur earlier as the massive star contracts to the main-sequence since winds should be launched before the star reaches the main-sequence. However, modeling the launching of UV-driven disk winds self-consistently for the simulations discussed in this work require extremely high resolution $(\sim R_{\rm \star}$) as demonstrated by \citet{Kee2019a} and therefore aren't currently computationally tractable. However, future work may be able to incorporate the results of  \citet{Kee2019a} as a disk wind sub-grid model in addition to the sub-grid wind model presented here. However, it remains unclear how efficient disk ablation is before the massive star reaches the main-sequence. 


\section{Conclusions}
\label{sec:conc}
In this work, we performed a series of 3D RMHD simulations of the gravitational collapse of isolated dense massive prestellar cores to determine how magnetic fields, turbulence, and stellar feedback from radiation pressure, collimated protostellar outflows, and isotropic radiatively-driven winds affects the formation of massive stellar systems. This is the first study of massive star formation to include stellar wind feedback, along with the stellar and dust-reprocessed radiation pressures and collimated outflows, in the context of a realistic, turbulent medium and self-consistent feedback evolution during star formation. By following the impact of stellar wind feedback we have investigated the production and expansion of adiabatic wind-driven bubbles in massive star formation with and without magnetic fields. 

We reach the following conclusions:
\begin{enumerate}
\item Radiatively-driven stellar winds are initially launched when massive stars are still actively accreting and contracting to the ZAMS. For the simulations presented here winds are initially launched when the star reaches $\sim$20 $\rm M_{\odot}$. 

\item The mass-loss rates and wind velocities evolve as the stars grow in mass and contract to the ZAMS. Therefore, protostellar evolution must be taken into account when modeling stellar wind feedback in massive star formation simulations.

\item We find that the kinetic energy and momentum injected by stellar winds is subdominant to that injected by stellar radiation and collimated outflows. Regardless, we find that winds are more likely to reduce the accretion flow onto massive stars when they are sufficiently massive.

\item We find that, for both magnetized and unmagnetized cores, the kinetic energy injected by stellar winds from massive stars produces hot shock-heated gas that expands adiabatically and launches expanding wind-driven bubbles or lobes that are asymmetrical and bipolar in morphology.  We name this phenomenon the ``wind tunnel effect," which is analogous to the ``flashlight effect" commonly seen in massive star formation simulations that only include radiative feedback. 

\item We find that the ``wind tunnel effect" occurs before the ``flashlight effect" when wind feedback is included. Like the ``flashlight effect," the hot gas produced by the shock-heating of stellar winds preferentially expands along regions where the density is lowest and therefore the presence of an accretion disk or dense circumstellar material pinches the expanding gas causing it to expand along the bipolar directions of the massive star.

\item By comparing identical simulations of the collapse of magnetized and unmagnetized prestellar cores that form massive stars we find that the wind-driven bubbles are launched at a lower stellar mass when magnetic fields are not included (at $\sim 27$~\msun\ versus $\sim 31$~\msun). We suggest that magnetic tension near the star delays the growth of these bubbles until stellar wind feedback is strong enough to overcome the magnetic tension thereby leading to the sustained expansion of adiabatic wind bubbles.

\item For unmagnetized cores we find that wind feedback can greatly inhibit the accretion of material onto massive stars once stellar winds are significant. In contrast, for magnetized cores we find that wind feedback initially enhances the growth rate of massive stars but once wind feedback becomes strong enough to launch wind-driven bubbles the accretion rate begins to decrease. Our results suggest that once stars become sufficiently massive, their strong stellar winds may inhibit accretion onto massive stars at late times.

\item By comparing the entrained wind and outflow material, we find that most of the wind material preferentially follows the low-density gas carved out by outflows. However, we find that the hot wind-driven bubbles are smaller in volume than the entrained wind material. In addition, we find evidence that mixing with the surrounding cooler gas occurs at the shells of the wind-driven bubbles.

\item Given that wind feedback appears to be effective at quenching accretion onto $\sim$30~\msun\ protostars, our results suggest that stars more massive than this likely form via larger-scale, high ram-pressure dynamical inflows from their host cloud to overcome wind feedback.

\item We showed that the diffuse ($\le 3$~keV) X-ray emission is highly attenuated due to the high column densities associated with the massive core and therefore is likely not detectable by \chandra. We do find that the hard ($\ge 3$~keV) X-ray emission may be detectable by \chandra\ in agreement with observations. However, future X-ray telescopes, with a higher sensitivity and better spatial resolution, may be able to study the soft X-ray emission produced by highly embedded massive protostars.
\end{enumerate}

\software{\textsc{yt} \citep{Turk2011a}, \orion\ \citep{Li2012a, Orion2021}, \harm\ \citep{Rosen2017a}, \chianti\ \citep{Dere1997a}}

\subsection*{Acknowledgements}
A.L.R. thanks the anonymous referee for their advice and suggestions which improved the manuscript. A.L.R. acknowledges support from NASA through Einstein Postdoctoral Fellowship grant number PF7-180166 awarded by the \textit{Chandra} X-ray Center, which is operated by the Smithsonian Astrophysical Observatory for NASA under contract NAS8-03060; and support from Harvard University through the ITC Postdoctoral Fellowship. A.L.R. would like to thank Grace Olivier, Laura Lopez, Stella Offner, Mark Krumholz, Mike Grudic, and David Guszejnov for insightful conversations regarding this work. A.L.R. would also like to thank her “supervisor,” Nova Rosen, for “insightful conversations” and unwavering support while this paper was being written. Her contributions are not sufficient to warrant co-authorship due to excessive napping and her lack of programming and writing skills.\footnote{Because she is a cat.} The simulations were run on the NASA supercomputer Pleiades located at NASA Ames. We use the yt package \citep{Turk2011a} to produce all the figures and quantitative analysis.
\bibliographystyle{apj}
\bibliography{../../refsALR/refs}
\end{document}